\documentclass[11pt,a4paper]{article}
\usepackage{jheppub}
\usepackage{amssymb}
\usepackage{dsfont}

\usepackage{dsfont,hyperref}

\numberwithin{equation}{section}

\newcommand{\reef}[1]{(\ref{#1})}
\newcommand{\be}{\begin{equation}}
\newcommand{\ee}{\end{equation}}
\newcommand{\bea}{\begin{eqnarray}}
\newcommand{\eea}{\end{eqnarray}}

\title{Bootstrapping the 3d Ising twist defect}

\author[a]{Davide Gaiotto}
\author[a]{Dalimil Mazac}
\author[b]{Miguel F. Paulos}

\affiliation[a]{Perimeter Institute,
Waterloo, Canada}
\affiliation[b]{Department of Physics,
Brown University,
Box 1843,
Providence, RI 02912-1843,
USA}

\abstract{Recent numerical results point to the existence of a conformally invariant twist defect in the critical 3d Ising model.
In this note we show that this fact is supported by both epsilon expansion and conformal bootstrap calculations. We find that our results are in good agreement with the numerical data. We also make new predictions for operator dimensions and OPE coefficients from the bootstrap approach. In the process we derive universal bounds on one-dimensional conformal field theories and conformal line defects.
}

\begin{document}
\maketitle
\section{Introduction}

Conformal field theories (CFTs) play an important role in many aspects of theoretical physics, from the concrete study of physical systems at criticality to abstract problems in mathematical physics. Although some CFTs can be given a weakly coupled description, the most interesting and commonly occurring CFTs are strongly coupled, and in three or more dimensions few analytic tool are available to study them. In supersymmetric or lower-dimensional examples, conformal invariant defects have played a useful role in probing the structure of CFTs. A conformal defect is a non-local observable, a modification of the theory which is localized on a lower-dimensional manifold and preserves an appropriate subgroup of the conformal group. It is natural to attempt to define and study defects in non-supersymmetric, commonly occurring CFTs. The 3d Ising model at criticality is a natural candidate: it is in a sense the simplest non-trivial 3d CFT and has been the subject of an intensive and rather successful analysis by a variety of theoretical and numerical tools \cite{Pelissetto2002a} such as the $\epsilon$-expansion and Monte Carlo simulations. More recently, interesting constraints on the Ising model \cite{ElShowk:2012ht} and other CFTs have been derived using the methods of the conformal bootstrap \cite{Beem:2013qxa,Alday:2013opa,Caracciolo:2009bx,ElShowk:2012hu,Fitzpatrick:2012yx,Gliozzi:2013ysa,Komargodski:2012ek,Kos2013,El-Showk:2013nia,Liendo2012,Maldacena:2011jn,Poland:2010wg,Poland:2011ey,Rattazzi:2010gj,Rattazzi:2010yc,Rattazzi:2008pe,Rychkov:2009ij,Vichi-thesis,Vichi:2011ux}.

The simplest possible conformal defect in a CFT is a boundary condition. Boundary conditions in the 3d Ising model have been the subject of some theoretical \cite{Diehl1981, McAvity1995} and numerical study \cite{Binder199017}. More recently, there have been attempts to ``bootstrap'' such boundary conditions \cite{Liendo2012}, by looking at two-point functions of bulk operators in the presence of the boundary. Another interesting example is a monodromy, or twist defect. The global $Z_2$ flavor symmetry of the Ising model allows for a natural definition of codimension two twist defects: under a rotation around the defect, local operators pick up a phase factor according to their $Z_2$ quantum numbers. Due to their topological nature, such defects are essentially guaranteed to flow to scale invariant defects in the IR and possibly to conformal defects.

Recently the authors of reference \cite{Billo:2013jda} have used Monte Carlo simulations to provide numerical evidence for the existence of a twist defect in the 3d Ising model. In this note, we aim to present further evidence for this from different points of view. We shall take a two-pronged approach: direct analytic calculations using $\epsilon$-expansion techniques; and the numerical methods of the conformal bootstrap. In both cases not only do we find excellent agreement with existent data, but we are also able to make new predictions that may be verified in the near future. As such, our work is a nice example of the interplay between theory, Monte Carlo simulations and the numerical bootstrap.

The $\epsilon$-expansion, introduced by Wilson and Fisher \cite{Wilson1972a}, provides a framework to study the critical $O(N)$ models in a perturbative setting. A drawback of this method is its disregard of the conformal symmetry, and another is that high accuracy requires computations to high loop orders and Borel resummation due to the asymptotic nature of the perturbative expansion \cite{LeGuillou1985}. Nevertheless, the $\epsilon$-expansion has been used to determine basic critical exponents in 3d rather precisely. The numerical bootstrap recently provided compelling evidence for the consistency of this method by identifying a family of solutions to crossing symmetry, interpolating between the 2d and 3d Ising model, and the 4d free scalar \cite{El-Showk:2013nia}. It is thus natural to use the $\epsilon$-expansion as a source of data on the twist defect in the 3d Ising model. Concretely, we will start with the twist defect in the free theory, add a $\phi^4$ coupling in the bulk and study correlation functions in the IR. The theory is expected to flow to the twist defect of the 3d Ising model. Performing one-loop computations, and setting $\epsilon=1$, we find good agreement with the Monte Carlo data. The one-loop deviation from the 3d free theory is always in the right direction, and often surprisingly close to the measured value. Note that defect scaling dimensions have been studied for Wilson lines in $3d$ $U(1)$ gauge theory with matter in \cite{kolezhuk2006theory}.

As was mentioned before, boundary conditions have been previously considered in the context of the conformal bootstrap. The main obstacle in such program is the lack of guaranteed positivity/unitarity constraints in the intermediate channel where the two bulk operators are fused together. Here we shall take a different approach, by considering directly correlators of defect operators. This guarantees positivity, but the price to pay is that it uses very little information about the bulk CFT itself, as the bulk operators do not appear in any fusion channel. The only properties of the bulk theory which affect directly the four-point functions on the defect are its symmetries. The 3d Ising model should be a reasonable candidate for such an analysis, because it is strongly constrained by its symmetries: in a sense, it is the simplest 3d CFT with a $Z_2$ flavor symmetry. It would be interesting to investigate if such a strategy may be successful in the study of boundary conditions (codimension one defects). In this paper, we focus on the codimension two twist line defects, and thus consider the conformal bootstrap in the one-dimensional world volume of the defect. 

The spectrum of operators on the defect contains operators of various $U(1)$ `spin' (corresponding to rotations around the defect), which can be integer of half-integer according to the $Z_2$ charge of the operator. Further, the spectrum should contain a protected ``displacement operator'' $D$, of spin $1$ and dimension $2$. This is the operator one would add to the defect Lagrangian to deform the defect away from a straight line. We shall consider four-point functions of the simplest local operator $\psi$ on the defect, the leading spin-$1/2$ operator, which occurs in the defect OPE of the $Z_2$-odd bulk field $\sigma$ (the Ising model spin field). However, in one dimension one must take care because a four-point function can be decomposed only into two crossing symmetry channels. There are therefore two crossing equations: in the four-point function $\langle \psi\bar \psi \psi \bar \psi \rangle$ both fusion channels have spin $0$; but the correlator $\langle \psi \psi \bar \psi \bar \psi\rangle$ has both spin $0$ and spin $1$ fusion channels. 

We shall explore the constraints following from the crossing equations, deriving universal bounds on one-dimension unitary CFTs. By forcing the spectrum to contain the displacement operator, we can derive a bound on the dimension of the leading parity-even spin-0 operator. In the extremal case where the bound is saturated, we can reconstruct a unique solution to crossing symmetry \cite{ElShowk:2012hu}, and we find that for a certain value of the OPE coefficient of $D$ the spectrum seems to match that of the defect, found both numerically and via $\epsilon$-expansion. We also obtain a number of other operator dimensions and OPE coefficients which can be thought of as specific predictions for future numerical tests.

%The second equation corresponds to 
%the four-point function $\langle \psi \psi \bar \psi \bar \psi\rangle$, and has a spin $1$ fusion channel. 

%At first we consider a single bootstrap equation, obtaining a universal bound on operator dimensions in unitary 1d CFTs. This equation, by itself, has no reference to the existence of a bulk theory. Perhaps unsurprisingly, we find that the bound is a featureless straight line, from which we cannot determine any information on the twist defect. By including crossing symmetry constraints from a different OPE channel, we include a basic constraint which identifies the defect as a codimension two object: the existence of a protected ``displacement operator'' $D$, of spin $1$ and dimension $2$. This is the operator one would add to the defect Lagrangian to deform the defect away from a straight line. We consider four-point functions of the simplest local operator $\psi$ on the defect, the leading spin $1/2$ operator which occurs in the defect OPE of $Z_2$-odd bulk operators. The standard bootstrap equation corresponds to the four-point function $\langle \psi\bar \psi \psi \bar \psi \rangle$, where all fusion channels have spin $0$. The second equation corresponds to 
%the four-point function $\langle \psi \psi \bar \psi \bar \psi\rangle$, and has a spin $1$ fusion channel. 

Here is a brief outline of this note. In section \ref{sec:defect}, we review the twist defect introduced in \cite{Billo:2013jda}. We work in the continuum limit, describing the expected symmetries, low-lying operators and the form of the operator product expansion. Section \ref{sec:epsexpansion} is concerned with $\epsilon$-expansion calculations. In section \ref{sec:bootstrap} we turn to the methods of the modern conformal bootstrap and conclude in Section \ref{sec:conclusions} together with suggestions for further research.

\section{The $Z_2$ Twist Defect} \label{sec:defect}
Let us recall \cite{Billo:2013jda} that the twist line defect in the 3d Ising model can be constructed on the lattice by flipping the Ising coupling on a semi-infinite half-plane ending on a line of the dual lattice. Such semi-infinite surface is a topological defect, since physics is invariant under its arbitrary deformations fixing the boundary line, provided we also flip the spins in between the original and deformed surface. The boundary of such topological surface defect is precisely a twist line defect. In the continuum limit, correlation functions become discontinuous (antiperiodic) across the surface. Presumably the same twist line defect lies at the IR end of the renormalization group flow from the free theory with a $Z_2$ twist defect generated by $\phi^4$ coupling in the bulk.

The global spacetime symmetry group of a $D$-dimensional Euclidean parity-invariant CFT is $O^+(1,D+1)$, where parity or sphere inversion switches between the two connected components. A conformal $ Z_2$ twist line defect thus breaks the bulk symmetry $O^{+}(1,4)\times  Z_2$ down to $O^{+}(1,2)\times O'(2)$, where $O'(2)$ is a double cover of the group of rotations and reflections fixing the defect, such that the rotation by $2\pi$ is identified with the nonidentity element of $ Z_2$. The dihedral symmetry $D_{8}$ of motions of the cubic lattice fixing the defect, discussed in \cite{Billo:2013jda}, is a subgroup of $O'(2)$. $O^+(1,2)$ is the spacetime symmetry group of the defect. At the level of Lie algebras, we have $so(1,2)=sl(2,\mathbb{R})$, and the connected components of $O^+(1,2)$ are switched by the reflection in a plane orthogonal to the defect or the sphere inversion centered on the defect. Following \cite{Billo:2013jda}, we call the former the $S$-parity.

In this note, we will be concerned with local operators living on the twist defect. In the Ising model, these correspond to local modifications of the lattice model in close proximity of the defect line. Applying radial quantization centered at a point on the defect, the defect local operators are seen to correspond to the states of the CFT quantized on a two-punctured sphere, with each puncture inducing the $ Z_2$ action on the bulk fields. The local operators fall into representations of the group $O^{+}(1,2)\times O'(2)$. The 1D conformal algebra $sl(2,\mathbb{R})$ is generated by operators $P,D,K$ (respectively translations, dilations and special conformal transformations) satisfying the commutation relations
\begin{equation}
[D,P]=i P\,,\quad [D,K]=-i K\,,\quad [K,P]=-2i D\,.
\end{equation} 
Physically relevant irreps are the highest-weight representations labelled by the scale dimension $\Delta\geq 0$ of the primary $\mathcal{O}(x)$, i.e. $[K,\mathcal{O}(0)] = 0$, $[D,\mathcal{O}(0)] = i\Delta\mathcal{O}(0)$. $\Delta<0$ would lead to correlation functions growing with distance and also violation of the unitarity bound by the first descendant.

The counterpart of unitarity in the Euclidean signature has been called `reflection-positivity'. In our setting, this property means that any correlation function of a configuration of real operators which is invariant under the $S$-parity is positive. Real operators in the Ising model are those appearing in the real operator algebra generated by the spin field. Reflection-positivity of the 3d Ising model is not spoiled by the defect line since the lattice transfer matrix in a plane perpendicular to the defect is unchanged with respect to the bulk theory. This leads us to define the (Euclidean) conjugate $C(\mathcal{O}(x))\equiv\bar{\mathcal{O}}(x)$ as complex conjugate composed with $S$-parity, so that $\langle\bar{\mathcal{O}}(x)\mathcal{O}(y)\rangle\geq 0$. $C$ is an antilinear map on the algebra of local operators which reverses the $O(2)$ spin and commutes with the other quantum numbers.

The commutativity properties of the symmetry algebra enable us to find a basis of defect primaries with well-defined $S$-parity, and $O(2)$ spin $s$, which is (half)integer for primaries even (odd) under the global $Z_2$. Each $s=0$ representation moreover carries $O(2)$-parity, denoted $B$. We are free to choose the phase of the $s=0$ primaries so that $C$ acts on them as the identity. The basis of $|s|>0$ primaries can be chosen so that $B(\mathcal{O}) = b_{\mathcal{O}}\bar{\mathcal{O}}$. From $BC = CB$ and $B^2=1$, we get $b_{\mathcal{O}} = e^{i\theta}$. Redefining $\mathcal{O}\rightarrow e^{-i\theta/2}\mathcal{O}$, we cancel the phase and get $B\mathcal{O} = \bar{\mathcal{O}}$, so that $|s|>0$ do not carry any $O(2)$-parity.

Exactly as in higher dimensions, conformal invariance fixes the form of two and three point functions. The difference in 1d is that the three point function coefficient $c_{\mathcal{O}_1\mathcal{O}_2\mathcal{O}_3}$ may depend on the cyclic order of the operators (signature of the permutation), since this order is invariant under the connected component of identity in the conformal group. In particular, note that for $x<y<z$
\begin{equation}
 \langle\mathcal{O}_1(x)\mathcal{O}_2(y)\mathcal{O}_3(z)\rangle = (-1)^{S_1+S_2+S_3}\langle\mathcal{O}_3(-z)\mathcal{O}_2(-y)\mathcal{O}_1(-x)\rangle\,,
\end{equation}
where $(-1)^{S_i}$ is the S-parity of $\mathcal{O}_i$. Hence
\begin{equation}
 c_{\mathcal{O}_1\mathcal{O}_2\mathcal{O}_3} = (-1)^{S_1+S_2+S_3}c_{\mathcal{O}_2\mathcal{O}_1\mathcal{O}_3}.
 \label{eq:opecoefs}
\end{equation}
Arbitrary cyclic permutations are generated by $P+K$. The sign in \eqref{eq:opecoefs} will play an important role in one of our bootstrap equations.

Primary operators on the defect satisfy the usual operator product expansion
\begin{equation}
 \mathcal{O}_1(x)\mathcal{O}_2(y) = \sum_{\mathcal{O}_3}\frac{c_{\mathcal{O}_1\mathcal{O}_2\bar{\mathcal{O}}_3}}{|x-y|^{\Delta_1+\Delta_2-\Delta_3}}\mathcal{D}_{\Delta_i}(x-y,\partial)\mathcal{O}_3(y),
\end{equation}
where the sum runs over defect primaries and
\begin{equation}
 \mathcal{D}_{\Delta_i}(x-y,\partial) = \sum_{n=0}^{\infty} \frac{(\Delta_1+\Delta_3-\Delta_2)_n}{n!(2\Delta_3)_n}(x-y)^n\partial^n
\end{equation}
is fixed by conformal symmetry. Moreover, bulk operators can be expanded in terms of the defect operators in the so-called bulk-defect OPE \cite{Cardy1990,McAvity1995} , which for a scalar primary in the bulk takes the form
\begin{equation}
 \phi(x,z,\bar{z})=\sum_{\mathcal{O}}C^{\phi}_{\mathcal{O}}\frac{\bar{z}^{s_{\mathcal{O}}}}{|z|^{\Delta_{\phi}-\Delta_{\mathcal{O}}+s_{\mathcal{O}}}}\mathcal{B}_{\Delta_{\mathcal{O}}}(|z|,\partial)\mathcal{O}(x)\,,\label{eq:bulkdefectope}
\end{equation}
where we use complex coordinates $z,\bar z$ for the transverse directions, the sum is over defect primaries, and $s_{\mathcal{O}}$ denotes the $O(2)$ spin of $\mathcal{O}$. Conformal symmetry in the presence of the defect fixes $\langle\phi(x,z,\bar{z})\bar{\mathcal{O}}(y)\rangle$ up to an overall constant $C^{\phi}_{\mathcal{O}}$, and consequently determines
\begin{equation}
 \mathcal{B}_{\Delta}(|z|,\partial) = \sum_{n=0}^{\infty}\frac{(-1)^n(\Delta)_n}{n!(2\Delta)_{2n}}|z|^{2n}\partial^{2n}\,.
\end{equation}
Notice that in particular, the boundary OPE coefficient $C_{\mathds 1}^{\mathcal \phi}$ gives the expectation value of $\phi$,
\bea
\langle \phi(x,z,\bar z)\rangle= \frac{C_{\mathbf 1}^{\mathcal \phi}}{|z|^{\Delta_\phi}}\,\label{eq:vev}
\eea
Applying a $2\pi$ rotation to \eqref{eq:bulkdefectope}, we see that the defect expansion of a bulk operator $\phi$ even (odd) under the global $Z_2$ contains only defect primaries with integer (half-integer) spins. Typically, the bulk-defect OPE will contain an infinite tower of defect primaries at each allowed spin. An exception is the bulk free field, studied below, which only features one defect primary at each spin.

The defect spectrum always contains the displacement operator $D(x)$ which, when added to the Lagrangian, generates deformations of the defect. Its dimension and quantum numbers are fixed by the Ward identity expressing the breaking of transverse translational symmetry by the defect
\begin{equation}
 \partial_a T^{ai}(x,z,\bar z) = D^{i}(x)\delta^2(z,\bar z),
\end{equation}
where $i$ label the transverse coordinates. Hence $\Delta_D = 2$, $s_D = 1$, and $D$ is even under $S$-parity.

Let us illustrate the above in the simplest setting -- the theory of the free massless real scalar $\phi$ in three dimensions, with twist defect for the global $Z_2$. Applying the bulk equations of motion to the bulk-defect OPE of $\phi$, we find that the scale dimension of the defect primary of (half-integer) spin $s$ appearing in the OPE is $\Delta_s = |s| + 1/2$. We will denote this tower of operators by $\psi_s$. 
The field $\phi$ (and consequently each $\psi_s$) is even under $S$-parity. Reality of $\phi$ implies $\psi_{-s} = \bar{\psi}_s$. The lowest-lying non-identity defect primary is $\psi\equiv\psi_{1/2}$ with scale dimension $\Delta_{\psi} = 1$. Since the scale dimension of the bulk spin field in the 3d Ising model is close to the free-field value, we expect the lowest-lying operator in the Ising defect spectrum to have dimension close to 1 and share the other quantum numbers with the free-theory $\psi$. Going back to the free theory, the $\bar\psi\psi$ OPE contains primary operators of schematic form $\mathcal{O}_n = \bar\psi\partial^n\psi$, $n\geq 0$. We have $\Delta_{\mathcal{O}_n} = n + 2$, $s_{\mathcal{O}_n} = 0$, and the $S$-parity, as well as $O(2)$-parity of $\mathcal{O}_n$ is $(-1)^n$.
The $\psi\psi$ OPE features primaries with schematic form $\mathcal{S}_n = \psi\partial^{2n}\psi$ for $n\geq 0$. This time, we obtain $\Delta_{\mathcal{S}_n} = 2n+2$, $s_{\mathcal{S}_n} = 1$, and the operators are even under $S$-parity. $\mathcal{S}_0$ is the only candidate for the displacement operator, since forming further OPEs will only create operators with dimensions greater than 2. In the next section, we will compute the first-order corrections to the scale dimensions of some of these operators, as well as their three point function constants at the Wilson-Fisher fixed point in $4-\epsilon$ dimensions. In particular, we will check that $D \equiv \mathcal{S}_0$ is indeed protected at this order.

\section{Epsilon Expansion} \label{sec:epsexpansion}

In order to study the properties of the twist defect at the Wilson-Fischer fixed point in $4-\epsilon$ dimensions, we start with the $D=2-\epsilon$ dimensional twist defect in the free theory and add a bulk $\phi^4$ interaction at the critical coupling. Since renormalization is a local property, the bulk flow is unaffected by the presence of the defect, and so the critical coupling is the usual $g=(4\pi)^{2}\epsilon/3 + O(\epsilon^2)$. Correlation functions of local bulk operators interpolate between two regimes -- when the typical distances between the insertions are much smaller than the distance from the defect, the correlation functions become those of the Wilson-Fisher fixed point with no defect. In the opposite case, the correlation functions are controlled by the CFT data of the defect. In the latter regime, the distance from the defect acts as a UV cutoff.

In this section, we use bulk perturbation theory to study bulk correlation functions in the defect regime and thus determine the data associated to some important defect operators to the first order in $\epsilon$. The reader uninterested in the details may skip directly to the results which are displayed in table \ref{tab:MCWF}.

\subsection{The two-point function in the free theory}
First, we will need the two-point function in the free theory alias the propagator. It is anti-periodic around the defect and satisfies
\begin{equation}
 -\nabla^{2}G_{0}(x_1,x_2) =\frac{4\pi^{D/2 + 1}}{\Gamma\left(\frac{D}{2}\right)} \delta^{D+2}(x_1-x_2),
 \label{eq:propdef} 
\end{equation}
where we chose the normalization standard in CFT literature, resulting in the asymptotics
\begin{equation}
 G_0(x_1,x_2) \stackrel{x_1\rightarrow x_2}{\sim}\frac{1}{|x_1-x_2|^{d}}\,.
 \label{eq:G0bulk}
\end{equation} 
Let $x$ denote coordinates in the whole space and $y$ those along the defect. The propagator can be easily found in momentum space
\begin{equation}
 G_{0}(x_1,x_2) = \frac{2\pi^{D/2}}{\Gamma\left( \frac{D}{2} \right)}\sum_{s\in\mathbb{Z}+\frac{1}{2}}\int\frac{d^{D}k}{(2\pi)^{D}} e^{is(\theta_1-\theta_2)}e^{ik\cdot (y_{1}-y_{2})}I_{|s|}\left(kr_{-}\right)K_{|s|}\left(kr_{+}\right),\label{eq:propagator}
\end{equation}
where the Fourier transform is over the coordinates along the defect, $\theta$ is the angle around the defect, $r_{-}=\min(r_1,r_2)$, $r_{+}=\max(r_1,r_2)$ and $I_s, K_s$ are the modified Bessel functions. The contribution from spin $s$ can be integrated to give
\begin{align}
 G_0(x_1,x_2,s) &= \frac{1}{4^{\Delta}}\frac{\Gamma\left(\Delta\right)}{\Gamma\left( \frac{D}{2} \right)\Gamma\left(\Delta -\frac{D}{2} + 1\right)}\frac{e^{is(\theta_1-\theta_2)}}{(r_1r_2)^{\frac{D}{2}}}\xi^{-\Delta}\times\nonumber\\
 &\times\phantom{}_2F_1\left(\Delta,\Delta - \frac{D}{2}+\frac{1}{2};2\Delta - D +1;-\frac{1}{\xi}\right),
 \label{eq:gs} 
\end{align}
where $\Delta = |s| + D/2$ is the scaling dimension of the primary field $\psi_s$ of spin $s$ induced on the defect by $\phi$ in the bulk, and
 \begin{equation}
  \xi = \frac{(y_1-y_2)^{2} + (r_1 - r_2)^{2}}{4 r_1 r_2}
 \end{equation}
 is one of the two conformal cross-ratios, the other being the relative angle. The computation can be simplified by using conformal invariance -- it is enough to evaluate the spin-$s$ propagator at $r_1=r_2$ since this fixes the dependence on $\xi$. $\xi\ll1$, $\Delta\theta\ll1$ is the regime controlled by the bulk CFT and $\xi\gg1$ the regime controlled by the defect data. Defect channel scalar conformal blocks for equal external dimensions can be read off from \eqref{eq:gs}, since these depend only on the internal dimension $\Delta$ and space-time dimension. To compute the properties of $\psi_s$, we will need the spin-$s$ two-point function in four dimensions, where \eqref{eq:gs} reduces to
 \begin{equation}
  G_0(x_1,x_2,s)\stackrel{D=2}{=} \frac{e^{is(\theta_1-\theta_2)}}{4r_1r_2}\frac{\xi^{-\frac{1}{2}}}{\sqrt{1+\xi}\left( \sqrt{\xi} + \sqrt{1+\xi} \right)^{2|s|}}.
  \label{eq:gs2} 
 \end{equation}
 We can check that the infinite sum over spins produces the correct short distance singularity. Indeed, the full free two-point function can be resummed for $\theta_1=\theta_2$
 \begin{equation}
  G_0(x_1,x_2) \stackrel{\theta_1=\theta_2}{=} \frac{1}{|x_1-x_2|^{D}}\frac{2\Gamma\left( \frac{D+1}{2} \right)}{\sqrt{\pi}\Gamma\left( \frac{D}{2} \right)}\xi^{-\frac{1}{2}}
  \phantom{}_2F_1\left(\frac{1}{2},\frac{D+1}{2};\frac{3}{2};-\frac{1}{\xi}\right).
  \label{eq:g2} 
 \end{equation}
 When $\xi\ll1$, this reduces to the expected \eqref{eq:G0bulk}.
  For completeness, let us note that the full two-point function can be found explicitly in $D=2$ by summing \eqref{eq:gs2}
 \begin{equation}
  G_0(x_1,x_2) \stackrel{D=2}{=} \frac{1}{|x_1-x_2|^{2}}\frac{\cos\left(\frac{\theta_1-\theta_2}{2}\right)}{\sqrt{1 + \xi}}.
  \label{eq:d2full}
 \end{equation}

\subsection{The two-point function at one loop}
\subsubsection{Leading defect operators of half-integer spin}
\begin{figure}[b]
  \centering
  \includegraphics[width=.5\textwidth]{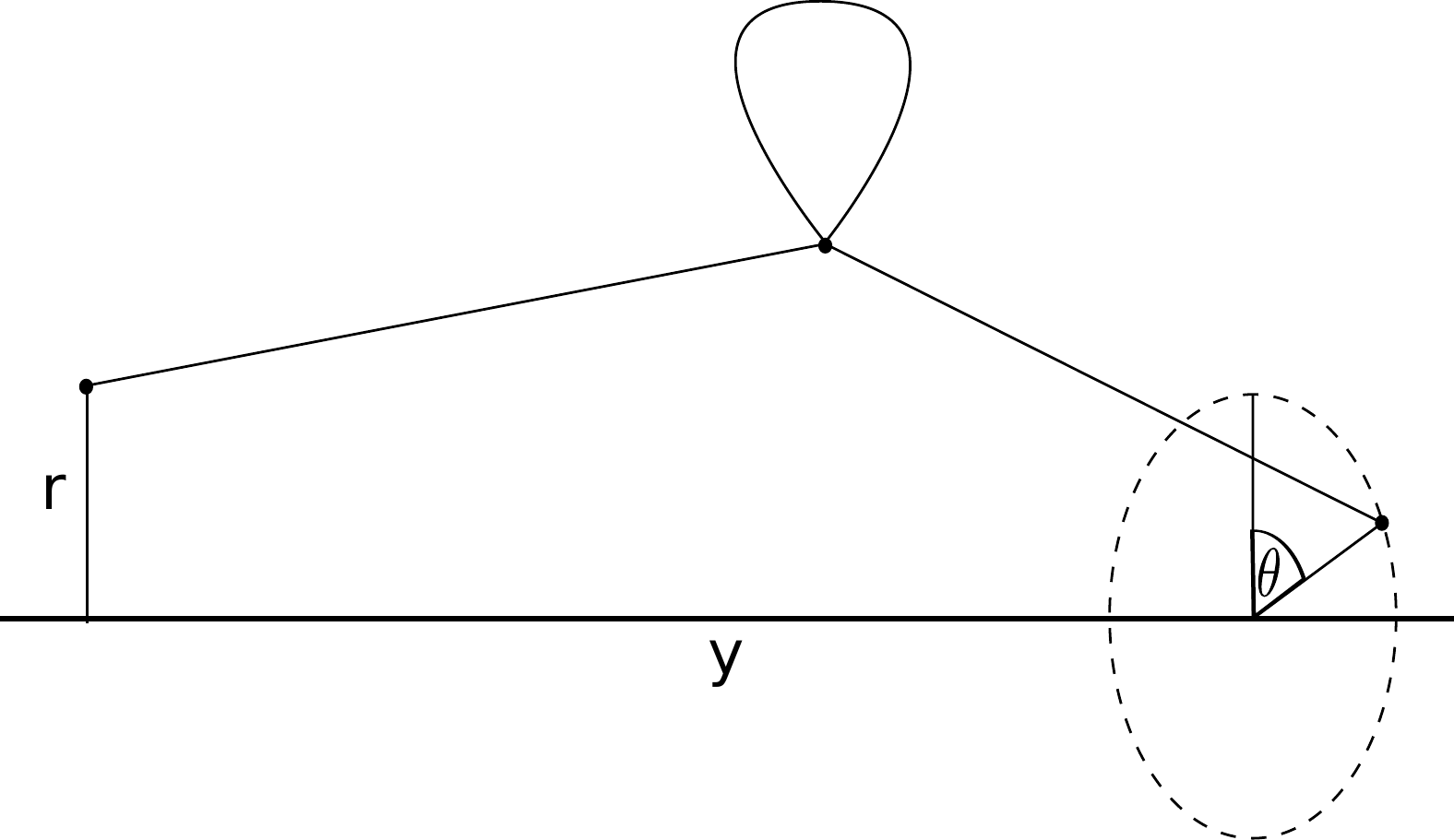}
  \caption{The one-loop contribution to $\langle\phi(x_1)\phi(x_2)\rangle$}
  \label{fig:2ptfunction}
  \end{figure}

In this subsection, we will compute the scaling dimensions of the operators $\psi_s$ of spin $s=n+1/2$, $n\in\mathbb{Z}_{\geq0}$, induced by $\sigma$ on the defect, as well as the bulk-defect OPE coefficient $C^{\sigma}_{\psi_s}$ to the first order in $\epsilon$. If nothing too dramatic happens along the RG flow from the free massless scalar, these should be the leading operators of half-integer spin. We will consider the spin-$s$ component of the bulk two-point function when the two insertions are taken close to the defect. Let us place both points at radius $r$ and distance $y$ along the defect, relative angle $\theta$ and denote $\lambda=r/y=1/\sqrt{4\xi}$. From the bulk-defect OPE, we expect the spin-$s$ component of the two-point function to have the following leading behaviour as $\lambda\rightarrow0$
\begin{equation}
  G(x_1,x_2,1/2) = |C^\sigma_{\psi_s}|^{2}\frac{e^{is\theta}}{r^{2\Delta_\sigma}}\lambda^{2\Delta_{\psi_s}}(1+O(\lambda^2)).
\end{equation}
The dependence of $\Delta_{\psi_s}$ and $C^{\sigma}_{\psi_s}$ on $\epsilon$ at one loop comes from two sources -- the change of the free theory result with space-time dimension and the one-loop self-energy diagram (see figure \ref{fig:2ptfunction}).
  Using \eqref{eq:gs}, one finds the free theory result
  \begin{equation}
    G_0(x_1,x_2,s) = \frac{\Gamma\left( s+ \frac{D}{2} \right)}{\Gamma\left(\frac{D}{2} \right)\Gamma\left( s+1 \right)}\frac{e^{is\theta}}{r^D}\lambda^{2s + D}(1+O(\lambda^2)).
  \end{equation}
  Expanding the Gamma functions, we obtain the free theory CFT data of $\psi_s$ to the first order in $\epsilon$
  \begin{align}
   \Delta_{\psi_s} &\stackrel{free}{=} s + 1 - \frac{\epsilon}{2}\\
   |C^{\sigma}_{\psi_s}| &\stackrel{free}{=} 1 + \frac{\psi(1)-\psi(s+1)}{4}\epsilon + O(\epsilon^2),
  \end{align}
  where $\psi(z)=(\log\Gamma(z))'$. The one-loop self-energy diagram should be evaluated in $D=2$ since the coupling constant is itself proportional to $\epsilon$. Taking care of the normalization and symmetry factor, the diagram's contribution is equal to
  \begin{equation}
    G_1(x_1,x_2,s) = -\frac{g}{32\pi^4}\int\limits_{\mathbb{R}^4} d^4 x_0\, G_0(x_1,x_0,s)G_0(x_0,x_0)G_0(x_0,x_2,s).
  \end{equation} 
  We need a regularized expression for the full free two-point function between coincident points $G_0(x_0,x_0)$ in $D=2$. Starting either from \eqref{eq:propagator} and evaluating the sum over spins for $D<0$ (so in dimensional regularization), or taking the finite piece of \eqref{eq:g2}, we find
  \begin{equation}
   G_0(x_0,x_0) = - \frac{\Gamma\left( \frac{D+1}{2} \right)}{2^{D-1}D\sqrt{\pi}\Gamma\left( \frac{D}{2} \right)}\frac{1}{r_0^D} \stackrel{D=2}{=}-\frac{1}{8}\frac{1}{r_0^2}.
   \label{eq:gxx} 
  \end{equation}
  Using $g = (4\pi)^2\epsilon/3$, and the free $D=2$, spin-$s$ propagator \eqref{eq:gs2}, and performing the trivial integration over the angle, the one-loop diagram becomes
  \begin{equation}
   G_1(x_1,x_2,s) = \frac{\epsilon}{24\pi}e^{is\theta}\int\limits_{\mathbb{R}^2}dy_0dz_0\int\limits_0^{\infty}\frac{dr_0}{r_0} \frac{(4r_0 r)^{2s}}{d_+ d_- e_+ e_- (d_+ + d_-)^{2s}(e_+ + e_-)^{2s}},
   \label{eq:gs1loop} 
  \end{equation}
  where
  \begin{align*}
   d_{\pm} &= \sqrt{\left( y_0 - \frac{y}{2} \right)^2 + z_0^{2} + \left( r_0\pm r \right)^2}\\
   e_{\pm} &= \sqrt{\left( y_0 + \frac{y}{2} \right)^2 + z_0^{2} + \left( r_0\pm r \right)^2}.
  \end{align*}
  When $\lambda\rightarrow 0$, the integral is proportional to $\lambda^{2(s+1)}\log\lambda$, which is giving precisely the anomalous dimension of $\psi_s$. The asymptotic expansion (see Appendix \ref{app:sintegral}) reveals that
  \begin{equation}
    G_1(x_1,x_2,s) =  - \frac{\epsilon}{12 s}\frac{e^{is\theta}}{r^2}\lambda^{2(s+1)}\left(\log\lambda + o(1)  \right)
  \end{equation}
  as $\lambda\rightarrow 0$. It follows that the one-loop contribution to $\Delta_{\psi_s}$ is $-\epsilon/24s$ and that to $|C^\sigma_{\psi_s}|$ vanishes. The CFT data at the Wilson-Fisher fixed point to the first order in $\epsilon$ are therefore
   \begin{align}
   \Delta_{\psi_s} &= s + 1 - \left( \frac{1}{2} + \frac{1}{24 s} \right)\epsilon + O(\epsilon^2)\label{eq:deltapsi} \\
   |C^{\sigma}_{\psi_s}| &= 1 + \frac{\psi(1)-\psi(s+1)}{4}\epsilon + O(\epsilon^2).
  \end{align}
  The inverse power-law dependence of the anomalous dimension on spin is in agreement with the results of \cite{Fitzpatrick:2012yx, Komargodski:2012ek}. The comparison to Monte Carlo data on $\psi=\psi_{1/2}$ and $\psi_{3/2}$ presented in \cite{Billo:2013jda} are reassuring, see table \ref{tab:MCWF}.
  
  \begin{center}
  \begin{table}\centering
   \begin{tabular}{c | c | c | l}
  \hline
  quantity & 3D free theory & Wilson-Fisher & Monte Carlo\\\hline
  $\Delta_{\psi}$ & 1 & 0.917 & 0.9187(6)\\
  $\Delta_{\psi_{3/2}}$ & 2 & 1.972 & 1.99(5)\\
  $\Delta_D$ & 2 & 2 & 2\\
  $\Delta_s$ & 2 & 2.167 & 2.27(1)\\
  $\Delta_{p^0}$ & 3 & 2.833 & 2.9(2)\\
  $\Delta_{t_+}$ & 3 & 3.111 & 3.1(5)\\
  $|C^\sigma_\psi|$ & 0.798 & 0.847 & 0.968(2)\\
  $|C^\sigma_{\psi_{3/2}}|$ & 0.651 & 0.680 & 0.61(9)\\
  $C^{\epsilon}_{\mathbf{1}}$ & -0.225 & -0.141 & -0.167(4)
  \end{tabular}
  \caption{A comparison of lattice data and the Wilson-Fisher fixed point at one loop}
  \label{tab:MCWF} 
  \end{table}
  \end{center} 
 
\subsection{Energy operator}
In this subsection, we will consider the two-point function in the bulk limit $\xi\ll1$ in order to find the one-point function of the energy operator $\epsilon$ in the presence of the defect at one loop. Put the two insertions at the same $\theta$, same radius $r$ and distance $y$ along the defect, so that $\lambda = r/y \gg 1$. Bulk OPE and conformal invariance of the one-point function dictates that
\begin{equation}
 G(x_1,x_2) = \frac{1}{y^{2\Delta_\sigma}}\left[1 + c_{\sigma\sigma\epsilon}C^{\epsilon}_{\mathbf{1}}\lambda^{-\Delta_\epsilon}(1 + o(1))\right],
\end{equation}
where the $o$-notation now refers to the limit $\lambda\rightarrow\infty$. Expanding the free-theory result \eqref{eq:g2} around $\xi = \infty$ yields
\begin{equation}
 G_0(x_1,x_2) =  \frac{1}{y^{D}}\left[1 -\frac{2^{-D}\Gamma\left( \frac{D+1}{2} \right)}{\sqrt{\pi}\Gamma\left( \frac{D+2}{2} \right)} \lambda^{-D}(1 + O(\lambda^{-2}))\right],
\end{equation}
which gives the following free-theory predictions for the CFT data associated to $\epsilon$
\begin{align}
 \Delta_{\epsilon} &\stackrel{free}{=} 2 - \epsilon\\
 c_{\sigma\sigma\epsilon}C^{\epsilon}_{\mathbf{1}} &\stackrel{free}{=} -\frac{1}{8}\left[1 + \frac{2\log2 - \psi(3/2) + \psi(2)}{2}\epsilon\right] + O(\epsilon^2).
\end{align}
The one-loop self-energy can be evaluated using the full 4D propagator \eqref{eq:d2full}. Rather than starting directly from \eqref{eq:d2full}, it is more convenient to sum \eqref{eq:gs1loop} over the spins, setting $\theta=0$
\begin{equation}
 G_1(x_1,x_2) = \frac{\epsilon}{3\pi}\int\limits_{\mathbb{R}^2}dy_0dz_0\int\limits_0^{\infty}dr_0 \frac{r}{d_+ d_- e_+ e_-}\frac{(d_++d_-)(e_++e_-)}{(d_++d_-)^{2}(e_++e_-)^{2}-(4rr_0)^{2}}.
 \label{eq:energy} 
\end{equation}
Asymptotic expansion of this integral as $\lambda\rightarrow\infty$ shows (see Appendix \ref{app:eintegral})
\begin{equation}
 G_1(x_1,x_2) =  \frac{\epsilon}{y^{2}}\lambda^{-2}\left[\frac{1}{24}\log\lambda + \frac{\log 2}{12} + o(1)\right].
\end{equation} 
We checked this result agrees with the computation which uses the full propagator \eqref{eq:d2full}. Combining the tree-level and one-loop result, we find the following properties of $\epsilon$ at one loop
\begin{align}
 \Delta_{\epsilon} &= 2 -\frac{2}{3}\epsilon + O(\epsilon^2)\\
 c_{\sigma\sigma\epsilon}C^{\epsilon}_{\mathbf{1}} &= -\frac{1}{8}\left[1 + \frac{2\log2 + 3\psi(2)-3\psi(3/2)}{6}\epsilon\right] + O(\epsilon^2).
\end{align}
The formula for $\Delta_\epsilon$ is in agreement with the standard result obtained using perturbation theory without the defect. We reproduce the computation in Appendix \ref{app:nodefect} in order to find the OPE coefficient $c_{\sigma\sigma\epsilon} = \sqrt{2}(1-\epsilon/6) + O(\epsilon^2)$. It follows that the one-point function coefficient of energy is
\begin{equation}
 C^{\epsilon}_{\mathbf{1}} = -\frac{1}{8\sqrt{2}}\left[1 + \frac{1 + 2\log2 + 3\psi(2)-3\psi(3/2)}{6}\epsilon\right] + O(\epsilon^2).
\end{equation}
As shown in table \ref{tab:MCWF}, the first order result is again in a good agreement with Monte Carlo data.

\subsection{The four-point function}
\subsubsection{Leading defect operators of positive integer spin}
Operators on the defect of integer spin can be found in the $\psi_{s_1}\psi_{s_2}$ OPEs. The most important of these is the displacement operator of spin one and protected dimension $D+1=3-\epsilon$. In the free theory, the normal ordered product $\psi_{s_1}\psi_{s_2}$ has scaling dimension $|s_1|+|s_2|+2 - \epsilon$. Consequently, the space of lowest-lying operators of positive integer spin $s$ is generated by all $\psi_{s_1}\psi_{s_2}$ with $s_1,s_2>0$ and $s_1+s_2 = s$. After flowing to the Wilson-Fisher fixed point, this degeneracy is lifted. Let us denote $\mathcal{O}_{s,m} \equiv \psi_{m-\frac{1}{2}}\psi_{s-m+\frac{1}{2}}$ for $m=1,\ldots,\lfloor\frac{s+1}{2}\rfloor$, with the exception $\mathcal{O}_{2k-1,k}\equiv\psi_{k-\frac{1}{2}}\psi_{k-\frac{1}{2}}/\sqrt{2}$, so that $\mathcal{O}_{s,m}$ is normalized in the free theory. At the Wilson-Fisher fixed point, the matrix of two-point functions of $\mathcal{O}_{s,m}$s is, to the first order in $\epsilon$,
\begin{equation}
 \langle\mathcal{O}_{s,m}(y_1)\bar{\mathcal{O}}_{s,n}(y_2)\rangle = \frac{1}{y_{12}^{2s+4 - 2\epsilon}}\left[ \delta_{mn}- 2\epsilon(\log y_{12})\Delta^s_{mn}  \right],
\end{equation}
where we ignored the possible corrections sub-leading in $y_{12}$. Denoting $\delta_s$ the minimal eigenvalue of $\Delta^s_{mn}$, the lowest dimension at spin $s\in\mathbb{Z}_{>0}$ is, to the first order in $\epsilon$
\begin{equation}
 \Delta_{s} = s + 2 + \epsilon (\delta_s - 1).
\end{equation}
In particular, if the displacement $D=\mathcal{O}_{1,1}$ is protected, we should have $\delta_1=\Delta^1_{11} = 0$.
 
\begin{figure}[t]
 \centering
 \includegraphics[width=.5\textwidth]{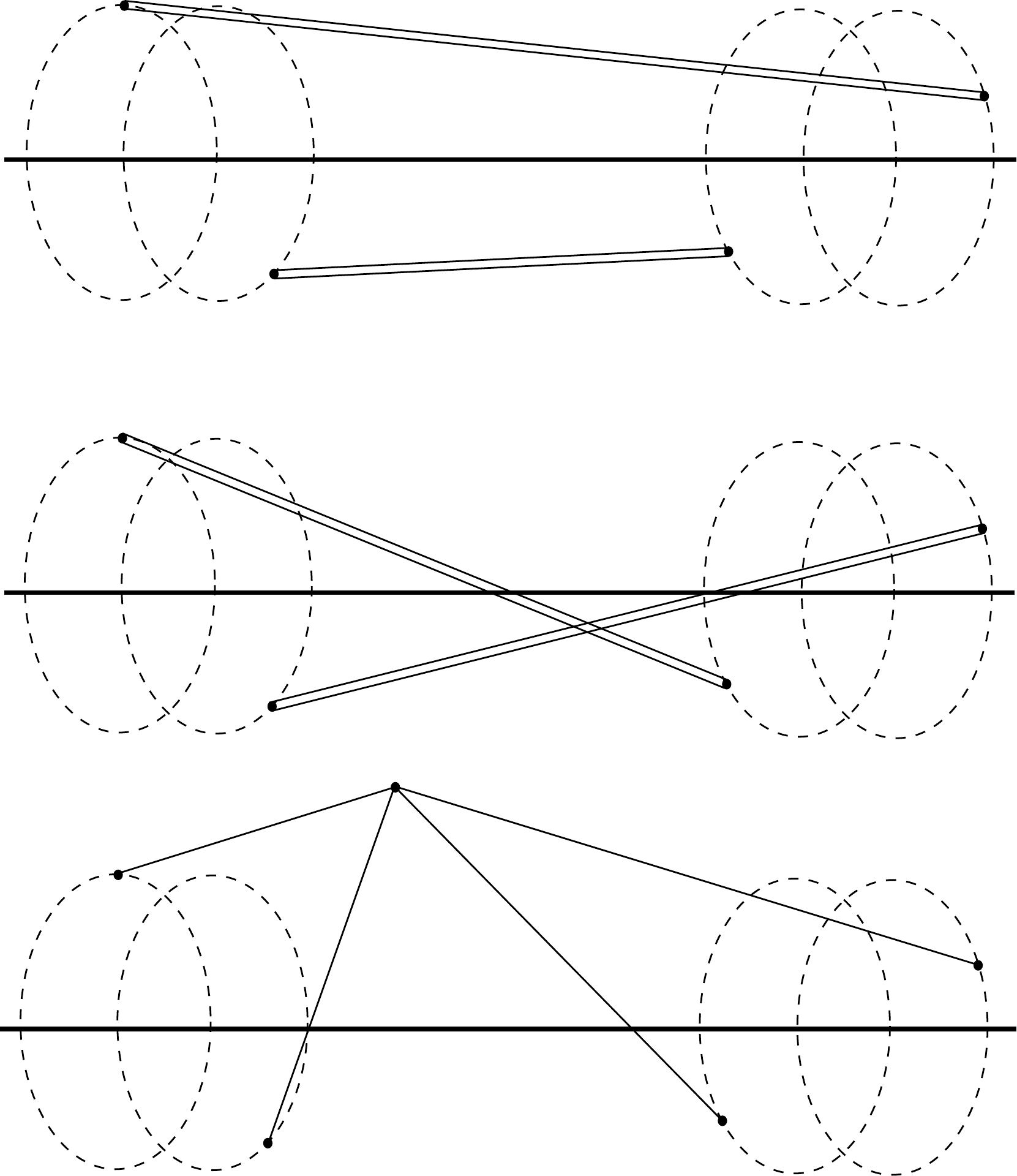}
 \caption{The diagrams contributing to the properties of $\psi_{s_1}\psi_{s_2}$ up to one loop. The double line denotes the one-loop-corrected propagator.}
\label{fig:4ptfunction}
\end{figure}

In the following, we will find the matrix $\Delta^s_{mn}$ by studying the various spin components of the four-point function of $\phi$ when all four insertions are at the same radius $r$ with $|y_{12}|=|y_{34}| = r/\lambda$ and $|y_{13}| = r/(\lambda\mu)$ such that $\lambda \ll 1$, $\mu\ll1$. Using first the bulk-defect OPE, and then OPE on the defect, we find the leading piece of the four-point function for $s_1,s_2>0$, $s_3,s_4<0$ and $s_1+s_2 = -s_3 - s_4 = s$
\begin{align}
  G\left(\{x_j,s_j\}_{j=1}^4\right) &= \frac{\prod_{j=1}^4\left(C^\sigma_{\psi_{s_j}} e^{is_j\theta_j}\lambda^{\Delta_{\psi_{s_j}}}\right)}{r^{4\Delta_{\sigma}}}\times\nonumber\\
  &\times c_{\psi_{s_1}\psi_{s_2}\bar{\mathcal{O}}_{s,m}}c_{\psi_{s_3}\psi_{s_4}\mathcal{O}_{s,n}}\mu^{2s+4-2\epsilon}\left[\delta_{mn} + 2\epsilon(\log\mu)\Delta^s_{mn}  \right],
 \label{eq:4ptfn} 
\end{align}
where $\mathcal{O}_{s,m}$ is the normalized product $\psi_{s_1}\psi_{s_2}$ and $\bar{\mathcal{O}}_{s,n}$ is the normalized product $\psi_{s_3}\psi_{s_4}$. Recall that to $O(\epsilon^0)$, we have $C^\sigma_{\psi_{s_j}} = 1$ and from Wick's theorem
\begin{equation}
 c_{\psi_{s_1}\psi_{s_2}\bar{\mathcal{O}}_{s,m}} =
 \begin{cases}
  1\quad&\textrm{if }s_1\neq s_2\\
  \sqrt{2}\quad&\textrm{if }s_1= s_2
 \end{cases}.
\end{equation}
In bulk perturbation theory, the contributions to the four-point function at the first order come from the diagrams with two disconnected loop-corrected propagators, and the contact four point interaction (see figure \ref{fig:4ptfunction}).
The former give the leading contribution
\begin{equation}
  G_{\mathrm{disc.}}\left(\{x_j,s_j\}_{j=1}^4\right) = \frac{\prod_{j=1}^4\left[ e^{is_j\theta_j}(\lambda\mu)^{\Delta_{\psi_{s_j}}}\right]}{r^{4\Delta_{\sigma}}}\left( \delta_{s_1,-s_3}\delta_{s_2,-s_4}+\delta_{s_1,-s_4}\delta_{s_2,-s_3} \right),
\end{equation}
while the contact interaction leads to the integral (following from \eqref{eq:gs2})
\begin{equation}
  G_{\mathrm{con.}}\left(\{x_j,s_j\}_{j=1}^4\right) = - \frac{\epsilon}{2^{7}3\pi}\int\limits_{\mathbb{R}^2}dy_0dz_0\int\limits_0^{\infty}\!\!\frac{dr_0}{r^4r_0^3}\prod_{j=1}^{4}\frac{e^{is_j\theta_j}}{\sqrt{\xi_j}\sqrt{1+\xi_j}\left( \sqrt{\xi_j} + \sqrt{1+\xi_j} \right)^{2|s_j|}},
 \label{eq:integerspindim}
\end{equation}
where
\begin{equation}
 \xi_j = \frac{(y_j - y_0)^2 + z_0 ^ 2 + (r-r_0)^2}{4rr_0}.
\end{equation} 
Asymptotic expansion gives the following leading piece (see Appendix \ref{app:Sintegral})
\begin{equation}
  G_{\mathrm{con.}}\left(\{x_j,s_j\}_{j=1}^4\right) = \frac{4\epsilon}{3(s + 1)}\left(\log\mu+O(1)\right)\frac{1}{r^4}\prod_{j=1}^4\left[e^{is_j\theta_j}(\lambda\mu)^{|s_j| + 1}\right],
 \label{eq:g4con} 
\end{equation}
which is consistent with \eqref{eq:4ptfn}. Putting the disconnected and contact interaction diagrams together, we find the following values of the matrix of scaling dimensions $\Delta^s_{mn}$
\begin{equation}
 \Delta^s_{mn} =
 \begin{cases}
  \frac{2}{3(s+1)}\quad&\textrm{if }m\neq n\\
  \frac{2}{3(s+1)}-\frac{1}{12}\left( \frac{1}{2m-1} + \frac{1}{2s-2m+1} \right)\quad&\textrm{if }m=n, 2m\neq s+ 1\\
  \frac{1}{3(s+1)} - \frac{1}{6s}\quad&\textrm{if }m=n, 2m= s+ 1
 \end{cases}
\end{equation}
The first term comes from the contact interaction and the second from the disconnected diagrams (if present), where we need to use the one-loop-corrected $\Delta_{\psi_{s_j}}$ from \eqref{eq:deltapsi}. The first case occurs when $\{s_1,s_2\}\neq\{-s_3,-s_4\}$, when only the contact interaction contributes. The second case occurs when $\{s_1,s_2\}=\{-s_3,-s_4\}$ but $s_1\neq s_2$. Finally, the third case occurs when $s_1=s_2=-s_3=-s_4$.
 
\begin{figure}[htb]
\centering
\includegraphics[width=.8\textwidth]{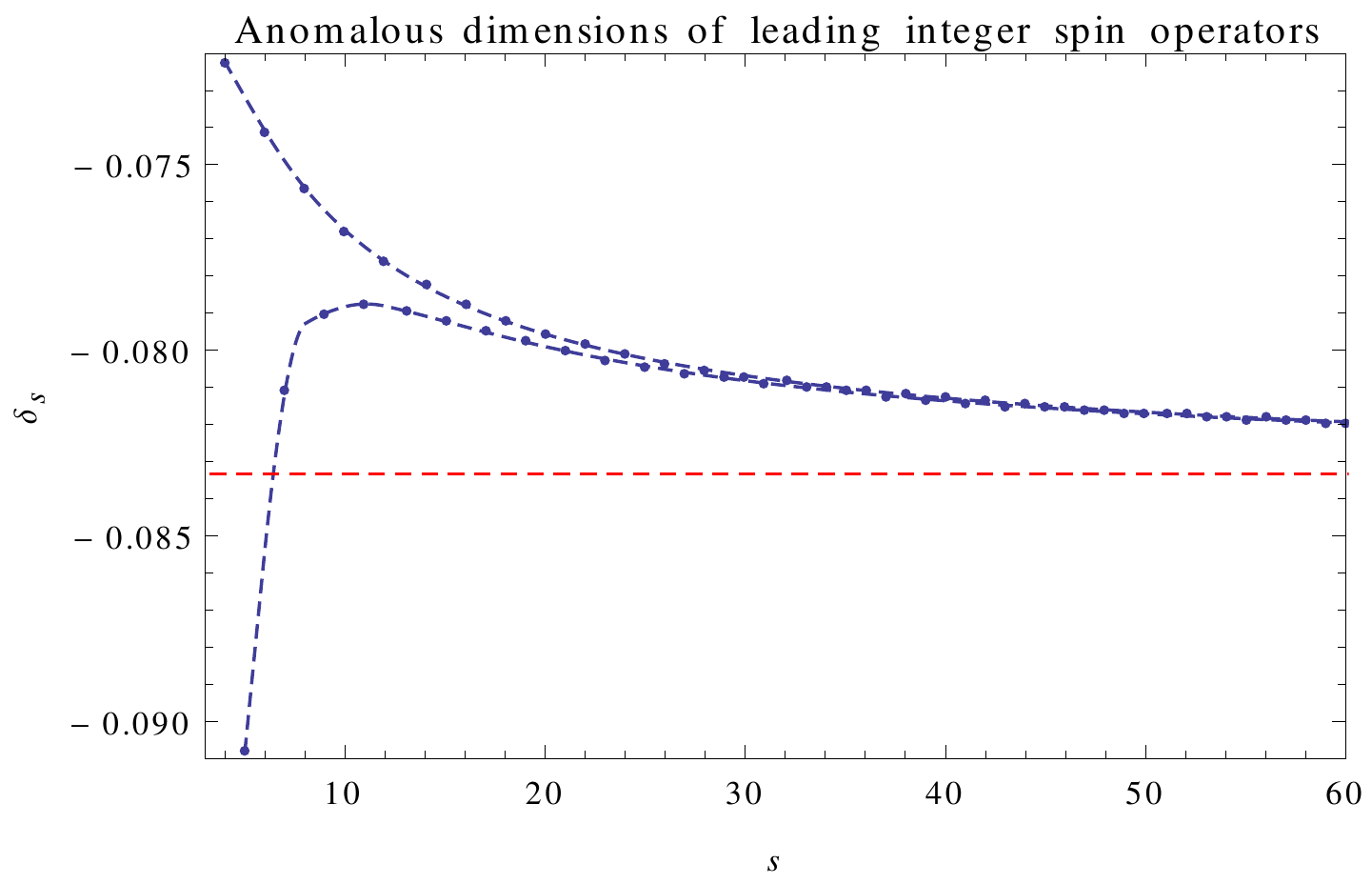}
\caption{Anomalous dimensions of the leading operators of spin $s$ at one loop. Dashed blue lines interpolate between the even and odd spins. They both asymptote to the dashed red line $\delta_s=-1/12$.}
\label{fig:anomdims} 
\end{figure}

The first thing to notice is that $\Delta^1_{11} = 0$, so the displacement operator is indeed protected at the first order in $\epsilon$. The next simplest case is $s=2$, with a single operator $t_{+} = \psi\psi_{\frac{3}{2}}$ of free-theory dimension $4-\epsilon$ and anomalous dimension $\epsilon/9$. Numerical results for the lowest eigenvalue of $\Delta^s_{mn}$ are shown in figure \ref{fig:anomdims}. The leading anomalous dimension converges to $-1/12$ as $s\rightarrow\infty$, which can be understood be noting that in this limit, $(e^{j})_n=\delta_{nj}$ becomes an eigenvector of $\Delta_{mn}^s$ with eigenvalue
\begin{equation}
 \lambda_j = -\frac{1}{12(2j-1)}\,.
\end{equation} 
It would be interesting to understand the asymptotic properties of the spectrum along the lines of \cite{Fitzpatrick:2012yx, Komargodski:2012ek}. Unfortunately, the Monte Carlo data on higher-spin operators are not yet precise enough to provide a test of our results.

Computation of the next-to-leading order in $\mu$ of the contact interaction diagram \eqref{eq:g4con} provides the first order correction to the OPE coefficients $c_{\psi_{s_1}\psi_{s_2}\mathcal{O}_{s,m}}$. The disconnected diagrams contribute only to $C^\sigma_{\psi_{s_j}}$. The computation is included in Appendix \ref{app:Dintegral}, the result being
\begin{align}
  G_{\mathrm{con.}}\left(x_1,\frac{1}{2};x_2,\frac{1}{2};x_3,-\frac{1}{2};x_4,-\frac{1}{2}\right) &= \epsilon\left( \frac{2}{3}\log\mu - \frac{8\log2 - 5}{6} + o(1)\right)\times\nonumber\\
  &\times\frac{(\lambda\mu)^6}{r^4}e^{(\theta_1+\theta_2-\theta_3-\theta_4)/2},
 \label{eq:4ptfnresult} 
\end{align}
from which it follows that
\begin{equation}
 c_{\psi\psi\bar{D}} = \sqrt{2}\left( 1 - \frac{8\log2 - 5}{24}\epsilon + O(\epsilon^2) \right).
\end{equation} 

\subsection{The leading defect scalar and pseudoscalar}
The above discussion was concerned only with operators of positive integer spin, but it is a simple matter to use the same method to find the dimension of the leading defect (non-identity) scalar. In the free theory, it is the operator $s=\bar{\psi}\psi$ of dimension $3-\epsilon$. Now we can repeat the steps above with $s_1 = - s_2 = -s_3 = s_4=1/2$ and find that the computation is almost identical to that for the displacement operator, the only difference being in the free-theory OPE coefficients ($c_{\psi\psi\bar{D}}=\sqrt{2}$, $c_{\psi\bar{\psi}s} = 1$). In both cases, the contribution from the disconnected diagrams is $-\epsilon/6$ (twice the anomalous dimension of $\psi$). The contact interaction diagram contributes $\epsilon/6$ to the displacement, but $\epsilon/3$ to the scalar since in the former case, it is reduced by $|c_{\psi\psi\bar{D}}|^2$=2. Hence the dimension of $s$ is
\begin{equation}
 \Delta_{s} = 3 - \frac{5}{6}\epsilon + O(\epsilon^2).
\end{equation}
Table \ref{tab:MCWF} indicates that already the first order provides a considerable improvement towards the Monte Carlo results with respect to the free theory. We can also use the constant piece of \eqref{eq:4ptfnresult} to conclude that
\begin{equation}
 c_{\bar\psi\psi s} = 1 - \frac{8\log2 - 5}{12}\epsilon + O(\epsilon^2).
\end{equation}

The leading free-theory defect operator with spin zero and negative $S$-parity is $p^0 = \bar\psi \overleftrightarrow\partial \psi/2 = [(\partial\bar\psi)\psi - \bar\psi(\partial\psi) ]/2$. We wish to study it using the $\langle \phi(x_1)\overleftrightarrow\partial\phi(x_2)\phi(x_3)\overleftrightarrow\partial\phi(x_4)\rangle$ bulk correlator, where the derivatives act along the defect. We put all four points at the same distance from the defect and focus on the correct spin component of the four-point function. The contact interaction diagram for $\langle\phi(x_1)\phi(x_2)\phi(x_3)\phi(x_4)\rangle$ is completely symmetric under any permutation of the four points. The antisymmetric derivative acting on $x_3$, $x_4$ thus makes the diagram vanish in the limit $x_3\rightarrow x_4$. Hence the properties of $p^0$ to the first order are determined solely by the renormalization of $\psi$. We find
\begin{equation}
 \Delta_{p^0} = 2\Delta_{\psi} + 1 + O(\epsilon^2) = 4 - \frac{7}{6}\epsilon + O(\epsilon^{2})\,.
\end{equation}
The generalized free theory gives for the three point function constant
\begin{equation}
 c_{\bar\psi\psi p^0} = \sqrt{\Delta_{p^0}} + O(\epsilon^2) = \sqrt{\frac{3}{2}}\left( 1 - \frac{7}{36}\epsilon + O(\epsilon^2) \right).
\end{equation}
We will be able to compare these predictions with data from conformal bootstrap in the following section.

\section{Bootstrapping the twist defect}\label{sec:bootstrap}

In this section we will apply the methods of the numerical conformal bootstrap to the one-dimensional defect directly. As outlined in the introduction, there are two distinct but related crossing equations which are relevant for our problem. Analysis of the first leads to an operator dimension bound in one dimension, similar to those derived between 2 and 4 dimensions in references \cite{El-Showk:2013nia,ElShowk:2012ht,Rattazzi:2008pe,Rychkov:2009ij}. The bound appears to be saturated by the generalized free fermion. Adding an extra equation and demanding the existence of a displacement operator leads to more interesting bounds, and we are able to reconstruct the twist defect spectrum. 

\subsection{The bootstrap equations}
The bootstrap equations that we use result from expanding four-point functions of $\psi$, $\bar\psi$ in different OPE channels. Four points on a line have only one invariant under the $SL(2,\mathbb{R})$ action. We take it to be
\begin{equation}
 z = \frac{x_{12}x_{34}}{x_{13}x_{24}}\,.
\end{equation}
We fix the order of the insertions to $x_1<x_2<x_3<x_4$, which results in the constraint $0<z<1$. The four-point function of defect primaries $\mathcal{O}_i$ of equal scale dimension $d$ can be written as
\begin{equation}
 \langle\mathcal{O}_1(x_1)\mathcal{O}_2(x_2)\mathcal{O}_3(x_3)\mathcal{O}_4(x_4)\rangle = \frac{1}{|x_{12}|^{2d}|x_{34}|^{2d}}g(z)\,,
\end{equation}
where $g(z)$ is an analytic function for $z\in(0,1)$. Colliding $x_1$ and $x_2$ leads to the series expansion in conformal blocks
\begin{equation}
 g(z) = \sum_{\mathcal{O}}c_{12\mathcal{O}}c_{34\bar{\mathcal{O}}} G_{\Delta_{\mathcal{O}}}(z),
\end{equation}
where the sum runs over defect primaries, and $G_{\Delta}(z)$ is the 1d conformal block for equal external dimensions and internal dimension $\Delta$. The conformal blocks are given by \cite{Dolan2011}
\begin{equation}
G_{\Delta}(z)= z^\Delta\, _2 F_1(\Delta,\Delta;2\Delta;z)\,.
\end{equation}
Colliding instead $x_2$ and $x_3$ and equating the two different representations of the four-point function leads to the crossing equation
\begin{equation}
  \sum_{\mathcal{O}}c_{12\mathcal{O}}c_{34\bar{\mathcal{O}}} z^{-2d}G_{\Delta_{\mathcal{O}}}(z) =  \sum_{\mathcal{O}}c_{23\mathcal{O}}c_{41\bar{\mathcal{O}}} (1-z)^{-2d}G_{\Delta_{\mathcal{O}}}(1-z)\,
\end{equation}
valid for $z\in(0,1)$.

$U(1)$ symmetry requires that a nonzero four-point function of $\psi$ and $\bar\psi$ must contain two of each. There are two nonequivalent orders to consider: $\langle\bar\psi\psi\bar\psi\psi\rangle$ and $\langle\bar\psi\psi\psi\bar\psi\rangle$. Focusing on the first case, the exchanged operators come from the $\bar\psi\psi$ OPE, so they have $U(1)$ spin zero. Moreover, their S-parity equals the $O(2)$ parity since the two symmetries require in turn
\begin{equation}
\langle \bar\psi\psi\mathcal{O}\rangle = (-1)^{S(\mathcal{O})}\langle\psi\bar\psi\mathcal{O}\rangle =(-1)^{B(\mathcal{O})}\langle\psi\bar\psi\mathcal{O}\rangle \,.
\end{equation}
We have seen this correlation between the parities in the $\bar\psi\psi$ OPE in the free theory example of section \ref{sec:defect}. Of course, the $\bar\psi\psi$ OPE starts with the identity. The coefficients of the conformal block expansion in the (12)(34) channel are $c_{\bar\psi\psi\mathcal{O}}c_{\bar{\mathcal{O}}\bar\psi\psi}$. Using the Hilbert space formalism, this equals
\begin{equation}
\langle\psi|\psi|\mathcal{O}\rangle\langle\mathcal{O}|\bar\psi|\psi\rangle = |\langle\psi|\psi|\mathcal{O}\rangle|^2 = |c_{\bar\psi\psi\mathcal{O}}|^2.
\end{equation}
The (23)(41) contains the same set of spin-0 operators and the corresponding coefficients are $|c_{\psi\bar\psi\mathcal{O}}|^2$. But $c_{\psi\bar\psi\mathcal{O}} = \pm c_{\bar\psi\psi\mathcal{O}}$ thanks to the parity symmetries, so that the first bootstrap equation can be written as
\begin{equation}
 \sum_{\mathcal{O}}|c_{\bar\psi\psi\mathcal{O}}|^2\left[z^{-2d}G_{\Delta_{\mathcal{O}}}(z)-(1-z)^{-2d}G_{\Delta_{\mathcal{O}}}(1-z)\right] = 0\,.
 \label{eq:beq1}
\end{equation}
We have thus obtained a conventional crossing equation with positive and equal coefficients on both sides, directly analogous to those used in higher dimensions \cite{ElShowk:2012ht,El-Showk:2013nia}.

The equation resulting from the crossing symmetry of the $\langle\bar\psi\psi\psi\bar\psi\rangle$ correlation function is less standard. The (12)(34) channel still consists of primaries from the $\bar\psi\psi$ OPE, but this time, the coefficient is
\begin{equation}
 c_{\bar\psi\psi\mathcal{O}}c_{\psi\bar\psi\bar{\mathcal{O}}} = 
 (-1)^{S(\mathcal{O})}c_{\bar\psi\psi\mathcal{O}}c_{\bar\psi\psi\bar{\mathcal{O}}} = (-1)^{S(\mathcal{O})}|c_{\bar\psi\psi\mathcal{O}}|^2,
\end{equation}
so that the conformal block expansion can distinguish between scalars and pseudoscalars at the cost of lost positivity. The (23)(41) channel comes from the $\psi\psi$ OPE, and so contains only spin-1 operators even under S-parity ($\langle\psi\psi\mathcal{S}\rangle =(-1)^{S(\mathcal{S})}\langle\psi\psi\mathcal{S}\rangle $). The coefficients are manifestly positive since
\begin{equation}
 c_{\bar{\psi}\bar\psi\mathcal{S}}c_{\psi\psi\bar{\mathcal{S}}} =
 \langle\psi|\bar{\psi}|\mathcal{S}\rangle\langle\mathcal{S}|\psi|\psi\rangle = |c_{\psi\psi\bar{\mathcal{S}}}|^2\,.
\end{equation}
The resulting bootstrap equation thus takes the form
\begin{align}
 \sum_{\mathcal{O}^+}|c_{\bar\psi\psi\mathcal{O}^+}|^2z^{-2d}G_{\Delta_{\mathcal{O}^+}}(z) &-
 \sum_{\mathcal{O}^-}|c_{\bar\psi\psi\mathcal{O}^-}|^2z^{-2d}G_{\Delta_{\mathcal{O}^-}}(z) =\nonumber\\
 &= \sum_{\mathcal{S}}|c_{\psi\psi\bar{\mathcal{S}}}|^2(1-z)^{-2d}G_{\Delta_{\mathcal{S}}}(1-z)\,,
 \label{eq:beq2}
\end{align}
where the first, second sum on the LHS runs over parity-even, odd scalars respectively, and the sum on the RHS runs over spin-1 primaries. We expect the lowest operator in the $\psi\psi$ OPE to be the displacement. Note that the difference in sign between the two bootstrap equations goes hand in hand with the fact that the crossed channel in \eqref{eq:beq1} starts with the identity, while in \eqref{eq:beq2}, it starts at $\Delta>0$. In the former case, the scalars and pseudoscalars together produce the strong singularity of the identity in the crossed channel, but in the later, their effect must cancel to leave a weaker singularity corresponding to the first spin-1 primary. Since the singularity in the crossed channel is produced by the tail of the set of primaries, it follows that there are infinitely many scalars as well as infinitely many pseudoscalars.

There is a family of simple solutions of the two bootstrap equations corresponding to a generalized free complex scalar in 1d.
In this case, Wick's theorem implies ($x_1<x_2<x_3<x_4$)
\begin{align}
 \langle\bar\psi(x_1)\psi(x_2)\bar\psi(x_3)\psi(x_4)\rangle &= \frac{1}{|x_{12}|^{2d}|x_{34}|^{2d}}\left[1 + \left( \frac{z}{1-z} \right)^{2d}\right]\label{eq:gf1}\\
 \langle\bar\psi(x_1)\psi(x_2)\psi(x_3)\bar\psi(x_4)\rangle &= \frac{1}{|x_{12}|^{2d}|x_{34}|^{2d}}\left(1 + z^{2d}\right)\label{eq:gf2}\,.
\end{align}
The first term in each bracket is the contribution of the identity, and the rest can be expanded in 1d conformal blocks as
\begin{align}
  \left( \frac{z}{1-z} \right)^{2d} &= \sum_{n=0}^\infty\frac{(2d)^2_n}{n!(4d+n-1)_n}G_{2d+n}(z)\label{eq:exp1}\\
  z^{2d} &= \sum_{n=0}^\infty\frac{(-1)^n(2d)^2_n}{n!(4d+n-1)_n}G_{2d+n}(z)\label{eq:exp2}\,,
\end{align}
so that the $\bar\psi\psi$ OPE contains scalars of dimensions $2d+2n$, $n\geq0$, and pseudoscalars of dimensions $2d+2n+1$, $n\geq 0$. \eqref{eq:gf2} in the crossed channel becomes
\begin{equation}
 \langle\psi(x_1)\psi(x_2)\bar\psi(x_3)\bar\psi(x_4)\rangle = \frac{1}{|x_{12}|^{2d}|x_{34}|^{2d}}\left[z^{2d}+\left( \frac{z}{1-z} \right)^{2d}\right]
\end{equation}
with conformal block expansion
\begin{equation}
 z^{2d}+\left( \frac{z}{1-z} \right)^{2d} = \sum_{m=0}^\infty\frac{2(2d)^2_{2m}}{(2m)!(4d+2m-1)_{2m}}G_{2d+2m}(z)\,,
\end{equation}
so that the spin-1 sector consists of dimensions $2d+2m$, $m\geq0$. 

Unless we put constraints on the spin-1 spectrum, any solution of \eqref{eq:beq1} can be extended to a solution of both \eqref{eq:beq1} and \eqref{eq:beq2}. Indeed, let
\begin{equation}
 \sum_{i}|\lambda_i|^2\left[z^{-2d}G_{\Delta_{i}}(z)-(1-z)^{-2d}G_{\Delta_{i}}(1-z)\right] = 0
\end{equation}
be a solution of the first equation and take the $\Delta>0$ spectrum in the even and odd scalar sectors identical, with $|c_{\bar\psi\psi\mathcal{O}^+_i}|^2 = |c_{\bar\psi\psi\mathcal{O}^-_i}|^2 = |\lambda_i|^2/2$. \eqref{eq:beq1} is automatically satisfied and in \eqref{eq:beq2}, the nonidentity scalars and pseudoscalars cancel out. Moreover, \eqref{eq:exp1} guarantees that we can use a tower of spin-1 operators of dimensions $2d + n$, $n\geq0$ to cancel the contribution of the identity.

Let us comment on the domain of applicability of our bootstrap equations. \eqref{eq:beq1} by itself does not know in any way about the bulk theory and merely expresses the constraints of crossing and unitarity for a 1d CFT. It is \eqref{eq:beq2} together with the assumption that the $\psi\psi$ OPE starts with the displacement that identifies the line as a codimension two object. Indeed, the structure of the OPE suggests a displacement operator which carries charge 1 under a transverse $SO(2)$ rotation symmetry, and a bosonic operator $\psi$ of half-integral rotation quantum number\footnote{Of course, the bounds derived from the bootstrap equations may apply to other situations which include operators with similar quantum numbers. For example, a codimension 3 defect has an $SO(3)$ 
rotation symmetry, and may have an operator of spin $1/2$ under that $SO(3)$. One could focus on a single component $\psi$ of that doublet and on the $SO(2)$ Cartan subgroup of the full rotation group, using our analysis for a sub-optimal bound.}.

\subsection{Constraints from the first crossing equation}

As a warm-up, let us consider first the constraints that follow from the first bootstrap equation \reef{eq:beq1}. This kind of equation has been previously analyzed in the literature, though not in one dimension. The major difference is that here there are no spin-$L$ representations other than $L=0$. Operators are  labeled only by their conformal dimensions, along with discrete quantum numbers. The method for deriving constraints from equation \reef{eq:beq1} has been explained in detail elsewhere, so here we will content ourselves with a brief summary. We first expand it in derivatives around $z=1/2$ up to some finite order. By setting each individual Taylor coefficient to zero, we are left with a system of linear equations with constraints, namely that the OPE coefficients should be positive and that at least one of them (that of the identity operator) is strictly non-zero. This is a linear programming problem, which can be solved with standard algorithms, such as the simplex method. Alternatively, we can try to disprove that such an equation can hold, by finding a linear functional which is non-negative on all possible vectors (namely, for any $\Delta$). We will follow the former route, using our own numerical implementation of the simplex algorithm. This has the advantage that the output is automatically a solution to the crossing symmetry constraints -- a spectrum, made up of operator dimensions and OPE coefficients, which solve the crossing equations -- as opposed to the linear functional method, where a spectrum has to be extracted by examining the zeros of the functional \cite{ElShowk:2012hu}. 

\begin{figure}[htb]
\begin{centering}
\begin{tabular}{cc}
\includegraphics[width=.47\textwidth]{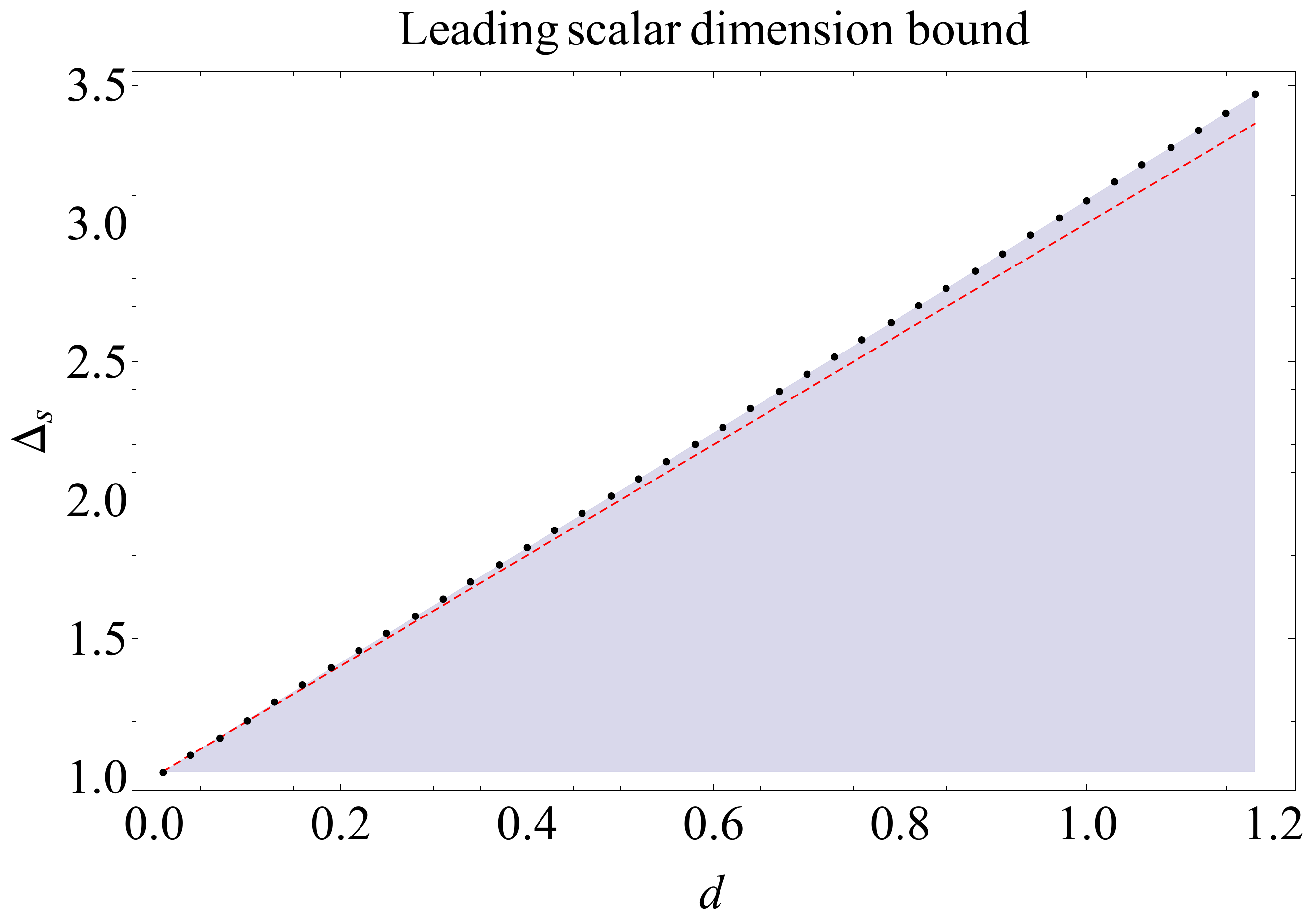} &
\includegraphics[width=.47\textwidth]{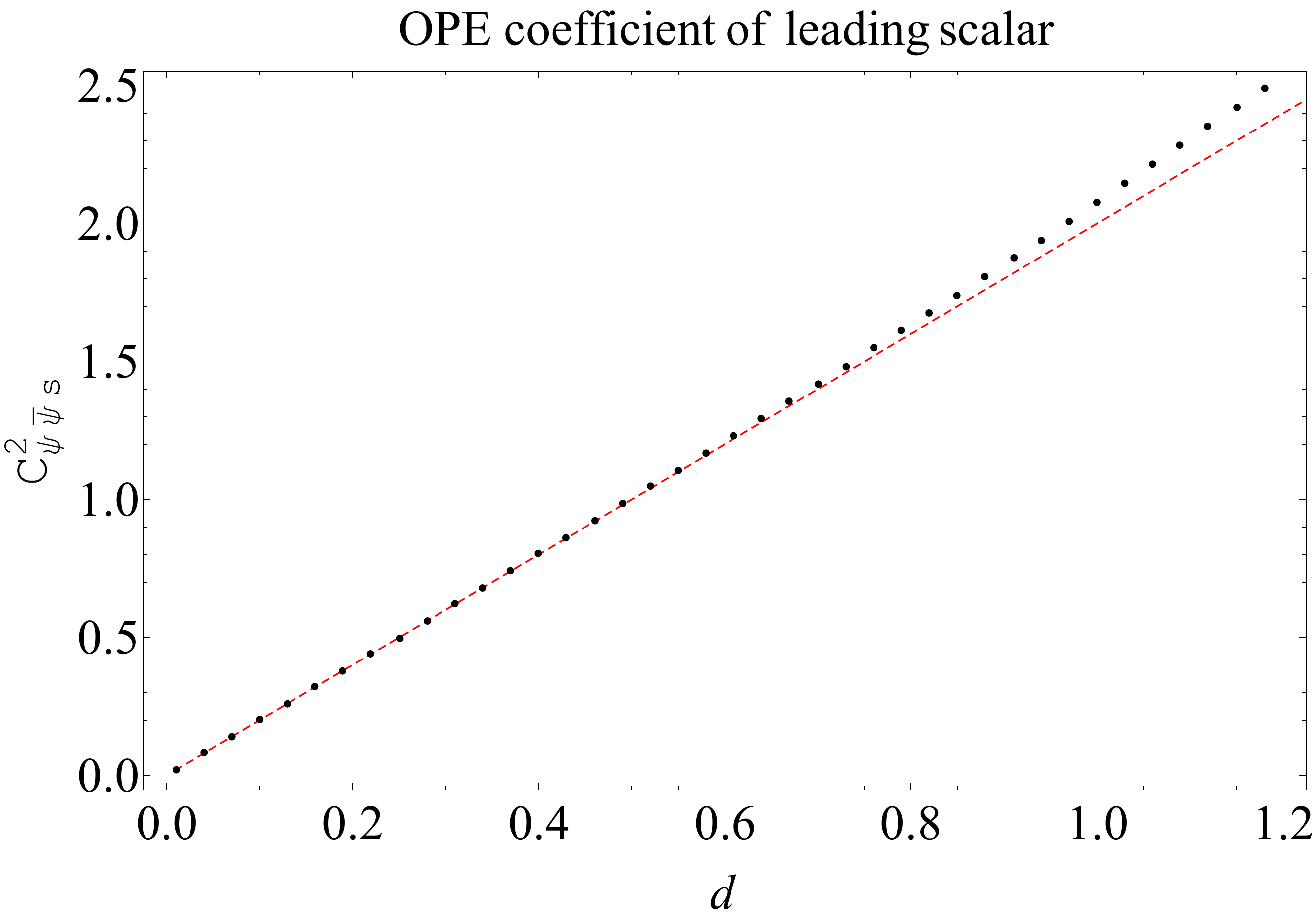} 
\end{tabular}
\caption{One-dimensional bounds derived from \eqref{eq:beq1}. In red the curves corresponding to the generalized free fermion solution. Left: bound on scalar dimension. Right: OPE coefficient of the leading scalar, in the solution to crossing corresponding to the dots on the top plot.}
\label{fig:1dBound1Eqn} 
\end{centering}
\end{figure}

Our approach is to fix $d$, the dimension of $\psi, \bar \psi$ and ask for the maximum allowed dimension of the first scalar appearing in the $\psi \bar \psi$ OPE. We do this by excluding from the sum rule \reef{eq:beq1} all vectors with dimension below some value $\Delta_s$ (apart from the identity). We then increase this gap until no solution can be found. The result is shown in figure \ref{fig:1dBound1Eqn}. 
%

%\begin{figure}[htb]
%\begin{centering}
%\includegraphics[width=.6\textwidth]{figures/1d_bound_dimension.pdf} 
%\includegraphics[width=.6\textwidth]{figures/1d_bound_ope_quad.pdf} 
%\caption{One-dimensional bounds. In red the curves corresponding to the generalized free fermion solution. Top: bound on scalar dimension. OPE coefficient of the leading scalar, in the solution to crossing corresponding to the dots on the top plot.}
%\label{fig:1dBound1Eqn} 
%\end{centering}
%\end{figure}
%

The result is a relatively boring straight line, which seems to very nearly coincide with the curve corresponding to the 1d generalized free fermion. This amounts to the four-point function
\begin{equation}
\langle\psi(x_1)\psi(x_2)\psi(x_3)\psi(x_4)\rangle = 
 \frac{1}{|x_{12}|^{2d}|x_{34}|^{2d}} \left[1+\left(\frac z{1-z}\right)^{2d}-z^{2d}\right]
\end{equation}
with conformal block expansion
\begin{equation}
1+\left(\frac z{1-z}\right)^{2d}-z^{2d} = 1 +  \sum_{j=0}^{\infty}\frac{2(2d)^{2}_{2j+1}}{(2j+1)!(4d+2j)_{2j+1}}G_{2d+2j+1}(z)\,,
\end{equation}
so that the minimal exchanged primary above the identity has $\Delta_s = 2d + 1$. We can find solutions to crossing at any point below our bound curve. In the extremal case where we sit directly on the bound itself, the solution is generically unique \cite{ElShowk:2012hu}. In this case we expect this solution to closely match the generalized free fermion. On the same figure on the right-hand side we compare the OPE coefficient of the leading scalar obtained with the bootstrap with that of the generalized free fermion -- namely $|c_{\psi \bar \psi \mathcal O}|^2=2d$. Overall the agreement is quite good for small $d$ and gradually gets worse as $d$ increases. As we increase the accuracy in our numerical procedure, by augmenting the total number of derivatives (here we have used 50), the agreement gets better and better for larger and larger values of $d$. As for the twist defect CFT, it lies well inside the bound, and as such, through bounds alone we cannot reach it, at least not with a single equation. This is unlike the situation described in \cite{El-Showk:2013nia}, where the Ising model lies on an interesting point (a kink) in the dimension bound. Here we are not as lucky and must work a bit harder to obtain an interesting result.

\subsection{Constraints from both crossing equations}
\begin{figure}[htb]
\centering
\includegraphics[width=.8\textwidth]{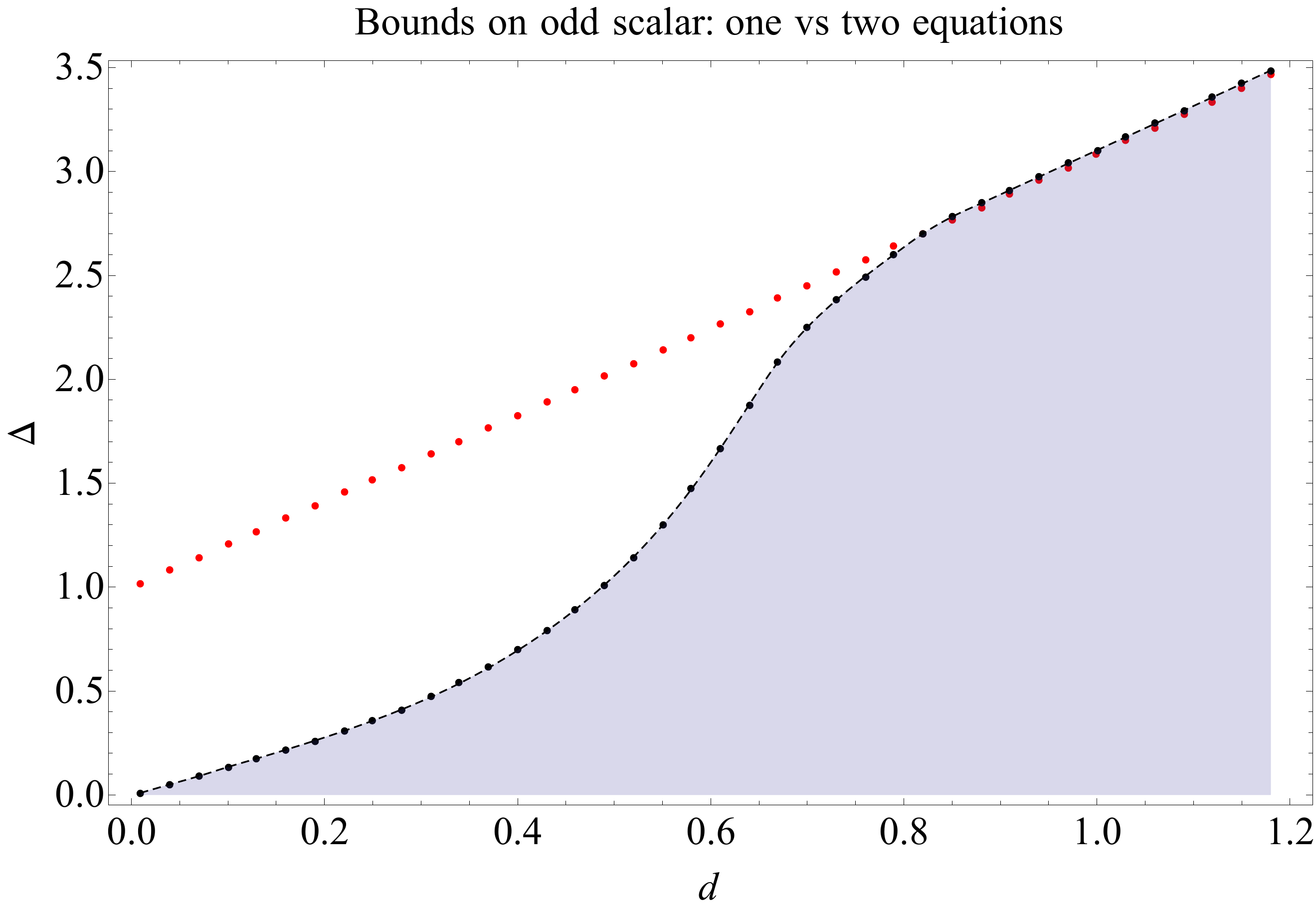}
\caption{Single equation bound in red and two equation bound in black. In the latter, the leading scalar is parity odd, up to about $d=1$, where the parity even and odd scalars have identical spectra.}
\label{fig:2eqnTemp} 
\end{figure}

\begin{figure}[htb]
\begin{centering}
\begin{tabular}{cc}
\includegraphics[width=.47\textwidth]{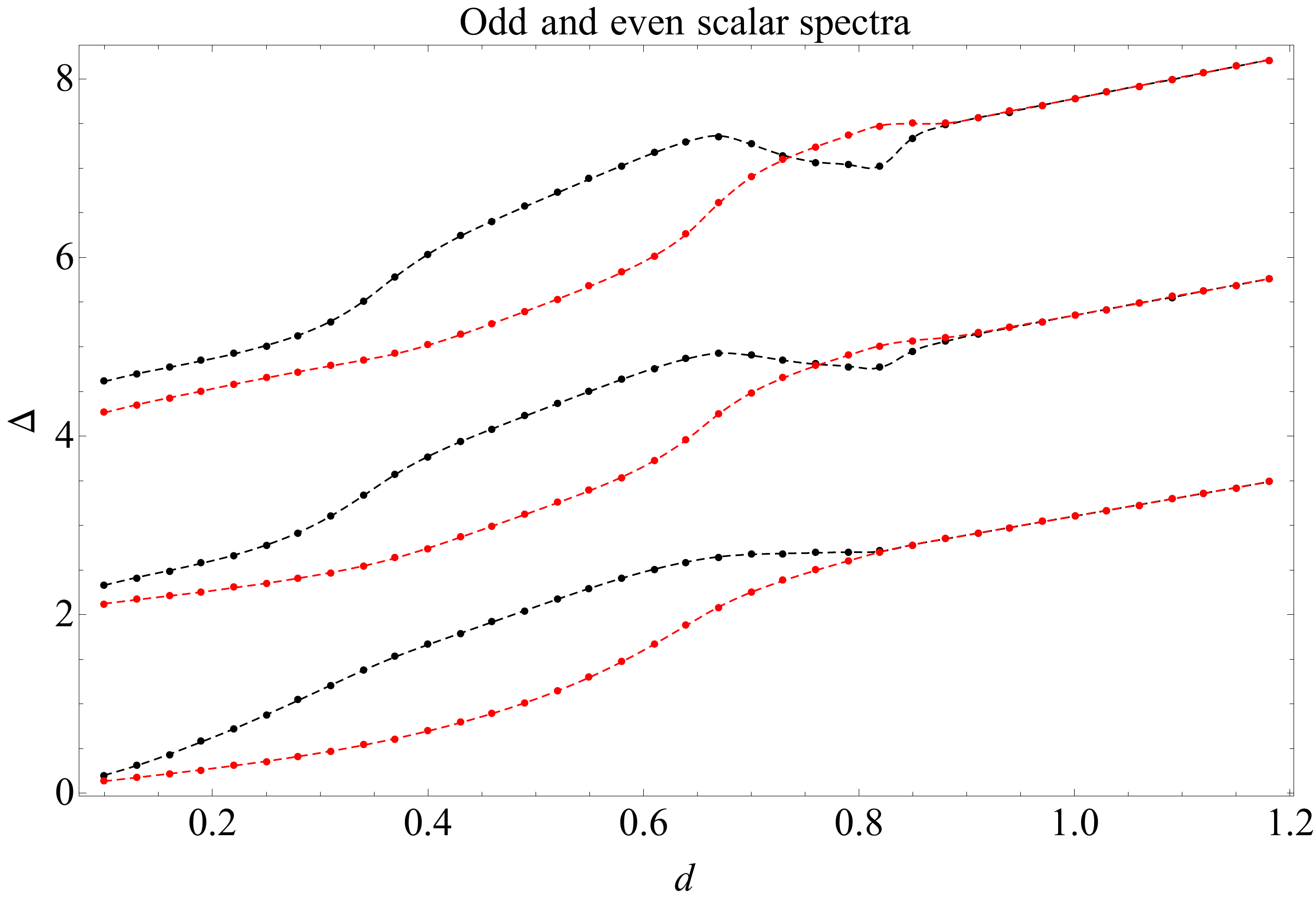} &
\includegraphics[width=.47\textwidth]{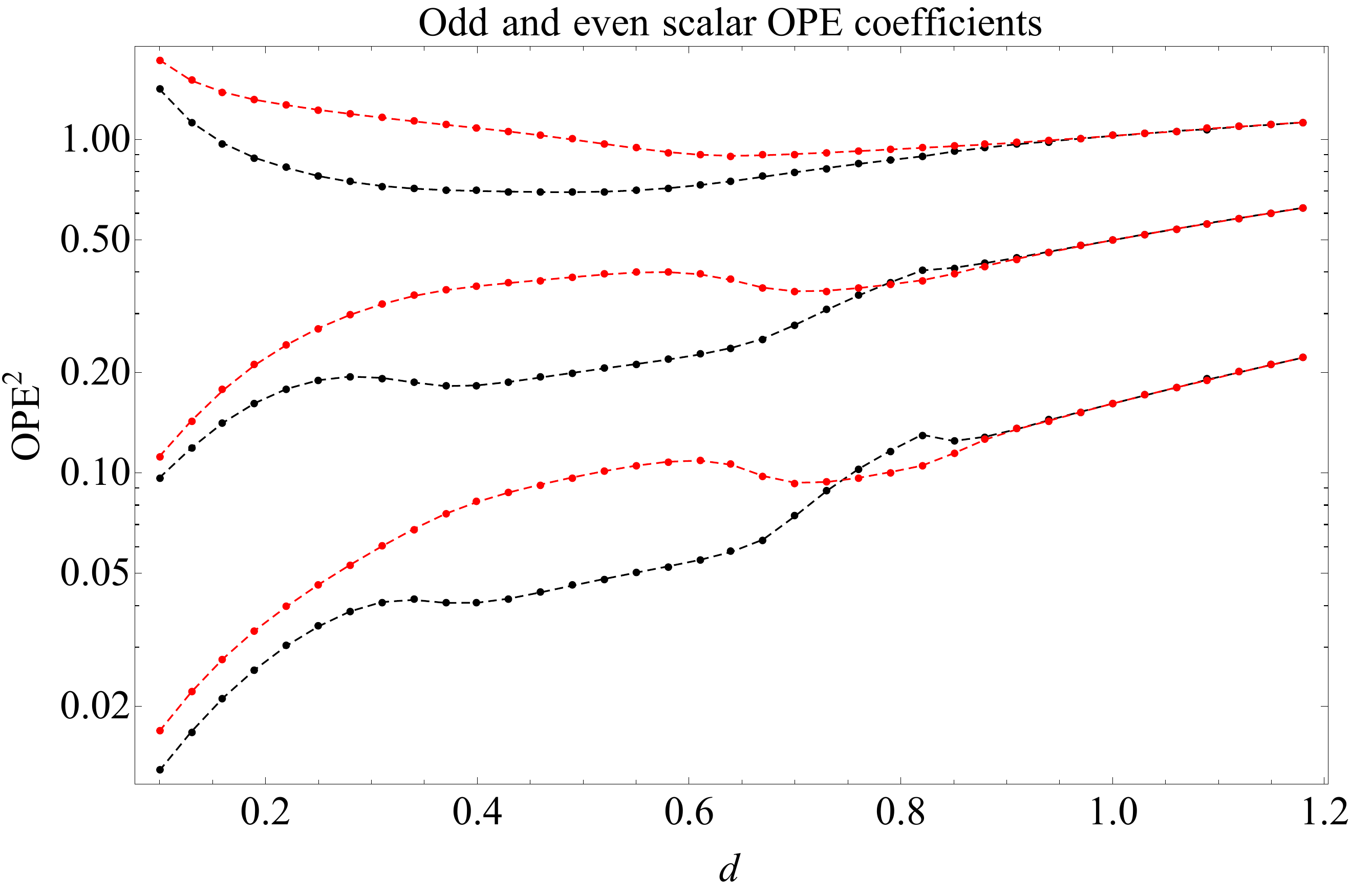} 
\end{tabular}
\caption{Spectra corresponding to the extremal solutions in figure \ref{fig:2eqnTemp}. In black (red) the parity even (odd) scalars. On the left the operator dimensions, and their OPE coefficients on the right. The correspondence between both is reversed: larger OPE coefficients correspond to lower dimension operators. 
}
\label{fig:2EqnSpectra} 
\end{centering}
\end{figure}

We now turn to deriving constraints by using both crossing equations. We use the conformal dimension to label operators, and define
\bea
F_\Delta(z)&=&G_{\Delta}(z)- \left( \frac{z}{1-z} \right)^{2d}G_{\Delta}(1-z),\\
S_\Delta(z)&=&G_{\Delta}(z),\\
T_\Delta(z)&=& -\left(\frac {z}{1-z}\right)^{2d}G_{\Delta}(1-z)
\eea
With this notation, it follows that we can
write \eqref{eq:beq1} and \eqref{eq:beq2} in vector form. 
\bea
\sum_{\mathcal {O}^+}a^+_{\Delta} 
\left(\begin{tabular}{c}
$F_\Delta(z)$ \\
$S_\Delta(z)$
\end{tabular}\right)
+
\sum_{\mathcal {O}^-} a^-_{\Delta} 
\left(\begin{tabular}{c}
$F_\Delta(z)$ \\
$-S_\Delta(z)$
\end{tabular}\right)
+
\sum_{\mathcal {S}} b_{\Delta} 
\left(\begin{tabular}{c}
$0$ \\
$T_\Delta(z)$
\end{tabular}\right)=0 \label{2eqnsumrule}
\eea
where all coefficients appearing in the above are explicitly positive. The procedure now is the same as in the single equation case. We evaluate the sum rule and its derivatives at $z=1/2$ (up to 40) and attempt to find a solution imposing various constraints. Since the spectrum is now split into three different sectors, we have more freedom in setting up the problem. Since we are looking for the twist defect, we are interested in solutions to crossing where the first spin-1 operator is the displacement, which has dimension 2. Therefore we shall impose a gap, by disallowing any spin-1 operators with dimension below 2 in the sum rule above. Figure \ref{fig:2eqnTemp} shows the bound derived by scanning over the dimension $d$ of $\psi$ while imposing the same gap on the dimension of the parity odd and parity even scalars.
The bound is clearly more restrictive up to some value of $d$, beyond which it returns to the original single equation result. This can be understood by recalling that a solution of the first equation can be extended to a solution of both as long as the gap imposed in the spin-1 sector does not exceed $2d$. We can see this directly by examining the spectra of the solutions to crossing living at the boundary of the bound. In figure \ref{fig:2EqnSpectra} we show the odd and even scalar spectra corresponding to these solutions. It is clear that for high enough $d$ the spectra become identical in these two channels, as we expect. A detailed examination of the OPE coefficients shows that this occurs precisely at $d=1$.

\begin{figure}[htb]
\begin{centering}
\begin{tabular}{cc}
\includegraphics[width=.45\textwidth]{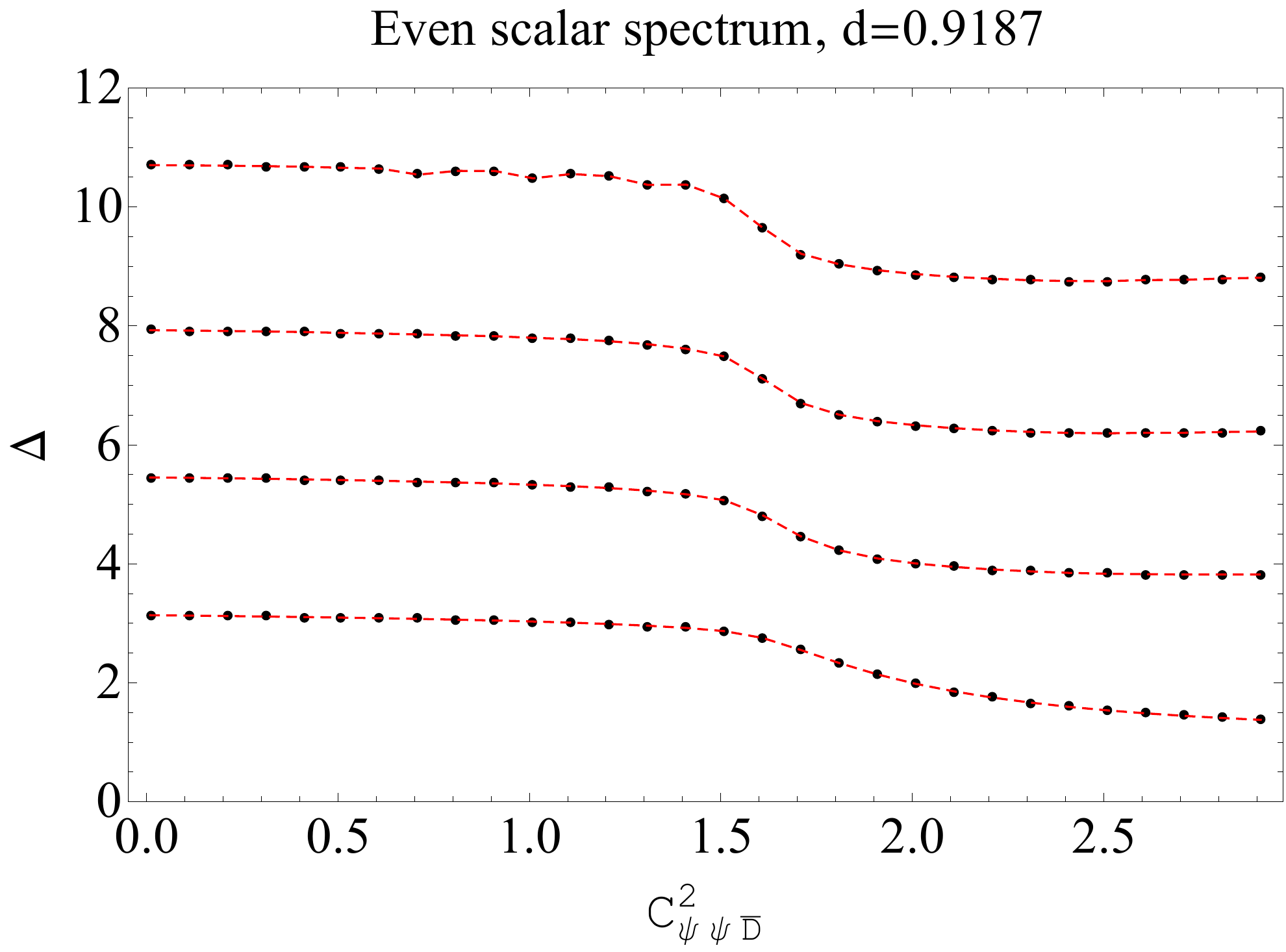} &
\includegraphics[width=.45\textwidth]{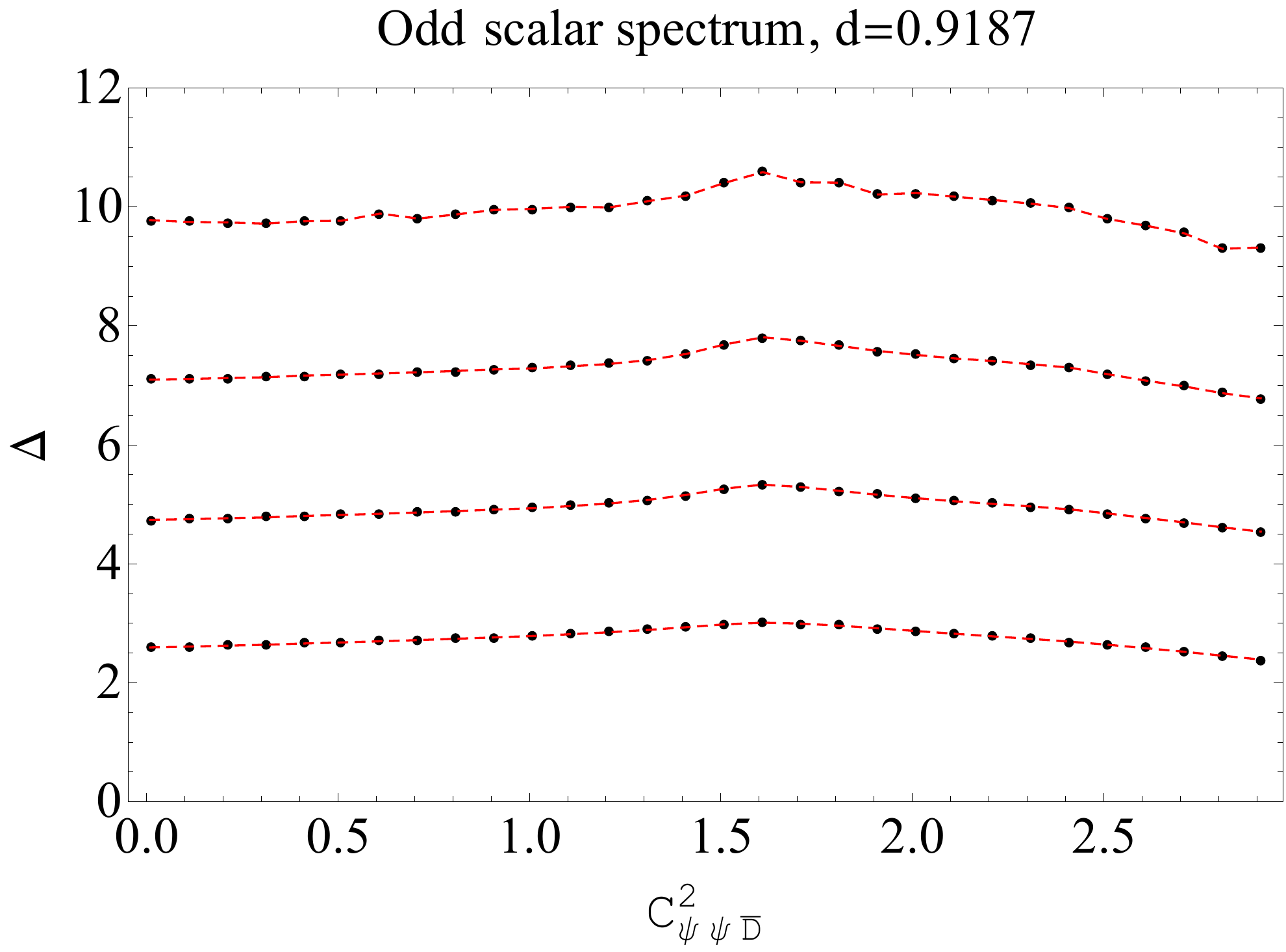} 
\end{tabular}
\includegraphics[width=.45\textwidth]{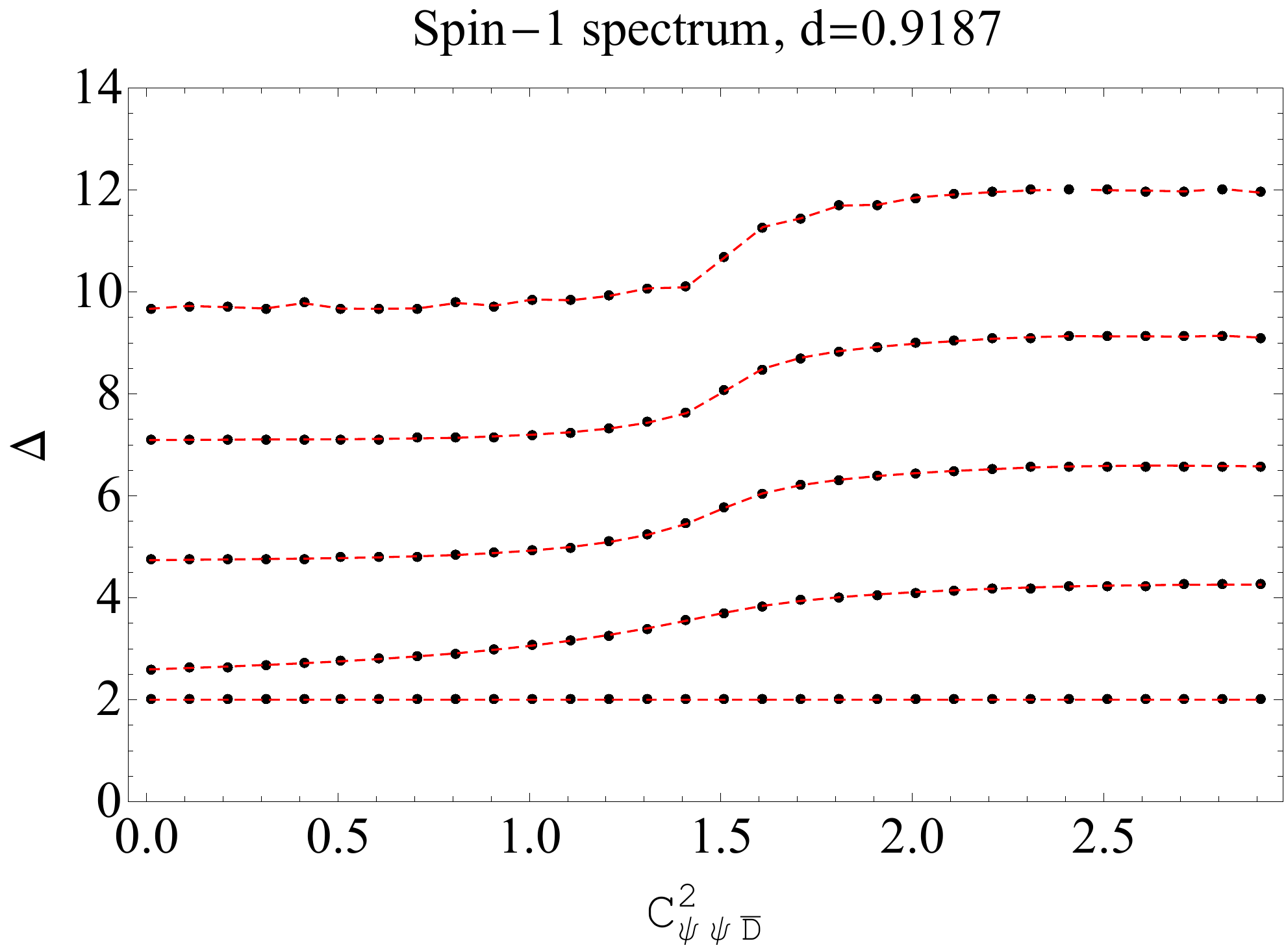} 
\caption{Spectra corresponding to the extremal solutions to crossing symmetry - the unique solutions at the boundary of our bounds.}
\label{fig:spectra} 
\end{centering}
\end{figure}

As it is clear, this approach is unfortunately still not sufficient to find the twist defect. From table \ref{tab:MCWF} we expect there to be a parity even scalar of dimension about 2.27 when $d\simeq 0.9187$, which is allowed, but not saturated by our bound. Hence we consider a different strategy. 
Since we know that the defect must contain a spin-1 operator with dimension 2 in its spectrum, we shall impose this directly on the sum rule. More concretely, we fix the OPE coefficient of the $D$ operator in the sum rule to some value, and we determine the maximum gap in the parity even sector consistent with crossing symmetry. 
We can do this for various values of $d$, but we will be interested in the experimentally relevant $d=0.9187$. 
\begin{figure}[htb]
\centering
\includegraphics[width=.7\textwidth]{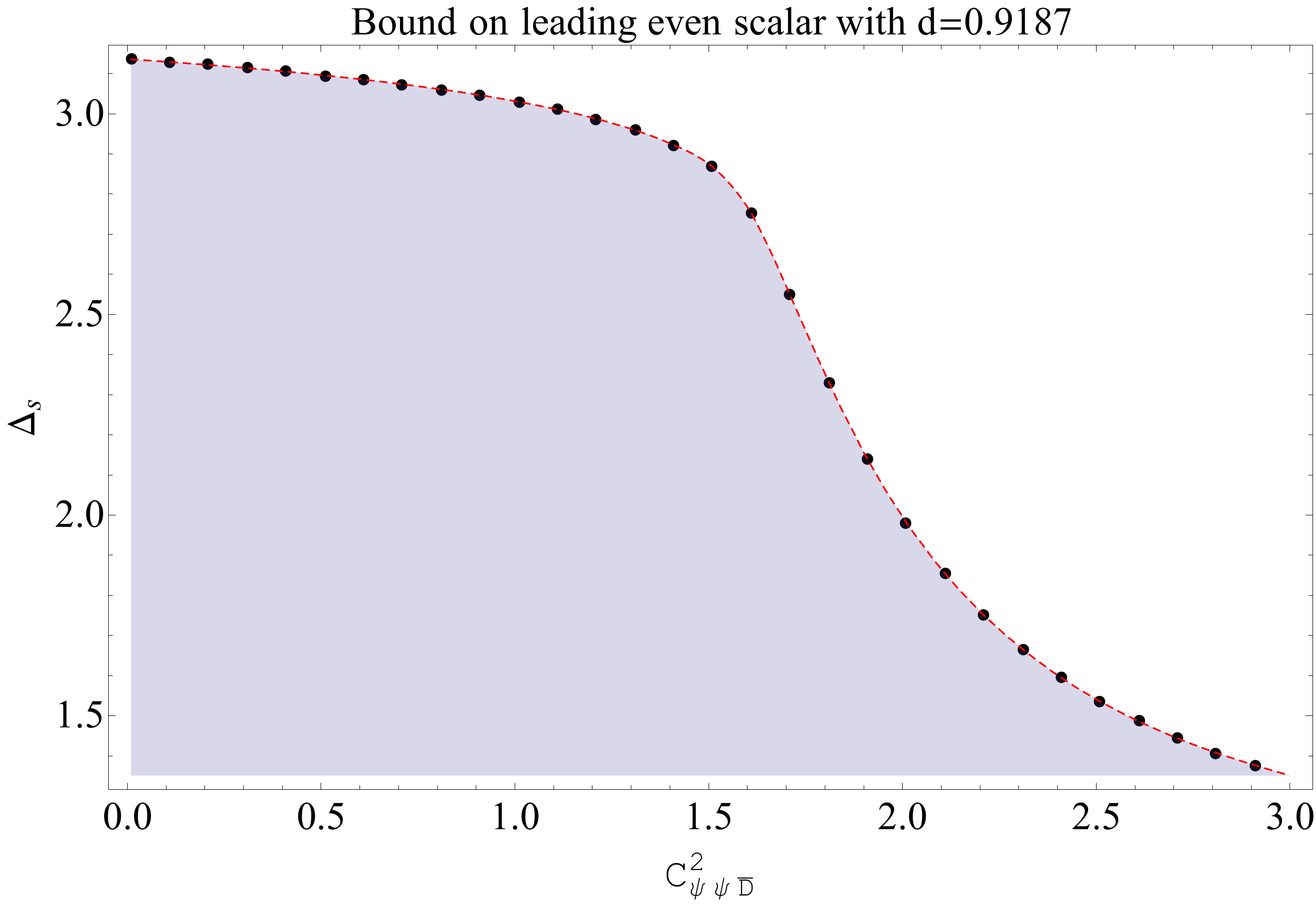}
\caption{One-dimensional bound, using two equations.}
\label{fig:2eqnOPEBound} 
\end{figure}
Figure \ref{fig:2eqnOPEBound} shows the resulting bound. We see that the bound is saturated by a solution to crossing including a parity even scalar of dimension $2.27$ for an OPE squared value of about $1.8$. Notice that this is consistent with the results of the $\epsilon$-expansion, which indicate that the OPE coefficient square should be $\simeq 1.9$. We can determine the spectrum of this solution, and this is shown in figure \ref{fig:spectra}. Remarkably, we find a parity odd scalar of dimension $\simeq 2.9$ in the solution, signaling that this is indeed the twist defect.
We summarize our spectrum results in table \ref{tab:Bootstrap}.
Besides the spectrum data present on the table, the bootstrap also predicts other operators and their OPE coefficients. The accuracy of these depends on the number of derivatives. We can estimate the error by repeating the calculations at different numbers of derivatives and seeing how the results change. Doing this we further predict the existence of the operators shown in table \ref{tab:Bootstrap2}, with an estimated error of $5\%$ or less.
\begin{center}
  \begin{table}\centering
   \begin{tabular}{c | c | c | c l}
  \hline
  quantity & Bootstrap & $\epsilon$-expansion & Monte Carlo\\\hline
  $\Delta_{\psi}$ & {\em 0.9187} & 0.917 & 0.9187(6)\\
  $\Delta_D$ & {\em 2} & 2 & 2\\
  $\Delta_s$ & {\em 2.27} & 2.167 & 2.27(1)\\
  $\Delta_{p^o}$ & 2.92 &      2.833      &  2.9(2)\\
    $c_{\psi\psi s}$ & 0.95 & 0.955 & ???\\
  $c_{\psi\psi\bar{D}}$ & 1.345 & 1.382 & ???\\
   $c_{\psi \bar{\psi} p^o}$ & 0.988 & 0.987 & ???
  \end{tabular}
  \caption{A comparison of lattice data, the Wilson-Fisher fixed point at one loop, and bootstrap calculations. We have italicized numbers which are used as inputs to the bootstrap method.}
  \label{tab:Bootstrap} 
  \end{table}
\end{center} 

To summarize, we have used as input the dimension of $\psi$; the dimension of the first even scalar $s$; and the existence of a spin 1 operator $D$ of dimension 2. Using this information, and assuming the defect spectrum lies on the bound of figure \ref{fig:2eqnOPEBound}, we have been able to determine the OPE coefficient of $D$ in the $\psi \psi$ operator product. Further, we have checked the existence of an odd scalar of dimension $\simeq 2.9$ and its OPE coefficient, and predict a further six operator dimensions and OPE coefficients. We could have gone further by doing more intensive calculations, but we are limited by the relatively large error in the dimension of $s$ determined from the lattice. As it stands, our confidence that we are finding the correct solution to crossing hinges on obtaining the correct operator dimension for $p^o$ and an OPE coefficient for the displacement operator consistent with the $\epsilon$-expansion. It would be very interesting to further test this by extending the $\epsilon$-expansion calculations or doing further lattice simulations.

\begin{center}
  \begin{table}\centering
   \begin{tabular}{c | c| c}
   \hline
   Type & Dimension & OPE$^2$\\\hline
   $0^+$ & 4.12 & 0.66 \\
   $0^+$ & 6.29 & 0.26\\
   $0^-$  & 5.11 & 0.45\\
   $0^-$  & 7.42 & 0.15\\
   1    & 3.98 & 0.99\\
   1    & 6.20 & 0.38
   \end{tabular}
  \caption{Spectrum predictions from the bootstrap method.}
  \label{tab:Bootstrap2} 
  \end{table}
 \end{center}

\section{Conclusions}\label{sec:conclusions}
We have offered new points of view on the twist line defect in the 3d Ising model -- the $\epsilon$-expansion and the conformal bootstrap of the defect four-point functions. While the $\epsilon$-expansion at one loop leads to a surprisingly good agreement with the Monte Carlo results, the identification of the defect spectrum from conformal bootstrap is not as straightforward as in the case of the bulk theory \cite{ElShowk:2012ht}. In spite of this, we believe we have successfully found the 1d defect theory by forcing the inclusion of the displacement operator in the spectrum, at the cost of using more data, namely the dimensions of the leading parity even scalar $s$ and of $\psi$ as determined from the lattice. The pay-off is that we determine a number of other quantities, namely operator dimensions and their OPE coefficients, which match well with results of the $\epsilon$-expansion. It is quite interesting that the inclusion of the second equation in the bootstrap set-up results in significant improvement of the bound, despite the lack of positivity in the spin-0 channel.

Several extensions of our work offer themselves. The $O(N)$ models allow twist line defects for arbitrary $R\in O(N)$, and it should be straightforward to generalize the $\epsilon$-expansion calculation at least in the case when $R=-I$. Although our bootstrap bounds apply to this defect for any $N$ by taking $\psi$ to be a fixed component of a spin-1/2 $O(N)$ vector, it may be worth repeating the analysis for $\langle\bar\psi_i\psi_j\bar\psi_k\psi_l\rangle$, $\langle\bar\psi_i\psi_j\psi_k\bar\psi_l\rangle$ while separating the exchanged primaries according to their $O(N)$ representations, as in \cite{Kos2013}. Large-$N$ computations for the defect should also be possible. Note that $O(N)$ can also be interpreted as the spacetime symmetry of the transverse directions, so that conformal bootstrap on the line can be used to constrain higher-dimensional CFTs. It may also be interesting to see how the bootstrap bound evolves for the $2-\epsilon$ dimensional defect in the Wilson-Fisher CFT.

1d CFTs can also serve as simple test cases for analytical understanding of the conformal bootstrap. In particular, the coincidence of the single equation bound with the generalized free fermion begs for an analytical explanation. Note that the techniques of \cite{Fitzpatrick:2012yx} and \cite{Komargodski:2012ek} are not directly applicable since they require the presence of two cross-ratios. Also for this reason, the study of crossing of the bulk two-point function in the presence of a defect may be a fruitful direction of research.

\acknowledgments
We are grateful for useful discussions with C. Beem, D. Simmons-Duffin, M. Meineri, R. Pellegrini, D. Poland, S. Rychkov, S. El-Showk and A. Vichi. The research of D.G. was supported by the Perimeter Institute for Theoretical Physics. Research at Perimeter Institute is supported by the Government of Canada through Industry Canada and by the Province of Ontario through the Ministry of Economic Development and Innovation. M.P. is supported by DOE grant DESC0010010-Task A. M.P. thanks CERN for hospitality while this work was being completed.

\appendix
\section{Asymptotic evaluation of integrals}
\subsection{Half-integer spin operators}\label{app:sintegral}
The one-loop properties of $\psi_s$ are encoded in the asymptotic properties of the integral \eqref{eq:gs1loop} as $\lambda\rightarrow0$. To find these asymptotics, we start by the substitution $y_0 = y a$, $z_0 = y b$, $r_0 = y c$, which leads to
\begin{equation}
  G_1(x_1,x_2,s) = \epsilon\frac{2^{4(s-1)}}{3\pi}\frac{e^{is\theta}}{r^2}\lambda^{2(s+1)}\int\limits_{\mathbb{R}^3}dadbdc\frac{c^{2s-1}}{d_+ d_- e_+ e_-(d_+ + d_-)^{2s}(e_+ + e_-)^{2s}},
\end{equation}
where
\begin{align*}
 d_\pm &= \sqrt{\left( a - \frac{1}{2} \right)^2 + b^{2} + \left( c\pm\lambda \right)^2}\\
 e_\pm &= \sqrt{\left( a + \frac{1}{2} \right)^2 + b^{2} + \left( c\pm\lambda \right)^2},
\end{align*}
and where we extended the domain of integration to the full $\mathbb{R}^3$, which is admissible since $2s-1$ is even. Let us denote
\begin{equation}
 I(\lambda) = \frac{2^{4(s-1)}}{3\pi}\int\limits_{\mathbb{R}^3}dadbdc\frac{c^{2s-1}}{d_+ d_- e_+ e_- (d_+ + d_-)^{2s}(e_+ + e_-)^{2s}}.
\end{equation} 
As $\lambda\rightarrow0$, the integral is logarithmically divergent around $(a,b,c) = \left( \pm 1/2 , 0 , 0 \right)$, with $\lambda$ acting as a point-splitting regulator. We expect $I(\lambda) = \alpha\log\lambda + \beta + o(1)$ and our goal is to determine $\alpha$ and $\beta$. Our general strategy will be to introduce an auxiliary parameter $N$ and split the integration domain into two parts. In this case, denote $I_1(\lambda,N)$ the integral above restricted to the union of the two spheres of radii $\lambda N$ surrounding the two singularities, and denote $I_2(\lambda,N)$ the integral over the rest of $\mathbb{R}^3$ so that $I(\lambda) = I_1(\lambda,N) + I_2(\lambda,N)$. $I_{1,2}(\lambda,N)$ simplify in the limit $N\rightarrow\infty$, $N\lambda\rightarrow 0$ if we do not care about terms which vanish as $\lambda\rightarrow0$. Working first with $I_1(\lambda,N)$, and focusing on the sphere surrounding $(a,b,c)=(1/2,0,0)$, note that in the limit $\lambda N\rightarrow 0$, we can replace $e_{\pm} = 1$. 
Making further the substitution $a = 1/2 + \lambda x$, $b = \lambda y$, $c =\lambda z$, we find that $\lambda$-dependence disappears
\begin{equation}
 I_1(\lambda,N) = \frac{2^{2s-3}}{3\pi}\int\limits_{x^2 + y^2 + z^2\leq N^2}\!\!\!\!\!dxdydz\frac{z^{2s - 1}}{f_+f_-\left(f_{+} + f_{-} \right)^{2s}} + o(1),
 \label{eq:I1} 
\end{equation}
where
\begin{equation}
f_{\pm} = \sqrt{x^2 + y^2 + (z\pm1)^2}.
\end{equation}
The integrals over $x$ and $y$ can be done explicitly, leaving us with
\begin{equation}
 I_1(\lambda,N) = \frac{2^{2s}}{12 s}\int\limits_{0}^{N}dz\left[ \frac{(z^{2s}+1)-|z^{2s}-1|}{2^{2s+1}z} - \frac{z^{2s-1}}{\left(\sqrt{N^2 + 2z +1}+\sqrt{N^2 - 2z +1}\right)^{2s}} \right] + o(1).
\end{equation}
It is now a matter of a simple calculation to show that
\begin{equation}
 I_{1}(\lambda,N) = \frac{1}{12 s}\log N + o(1).
\end{equation}
Let us consider $I_2(\lambda,N)$, denoting $D=\{(a,b,c)\in\mathbb{R}^3|(a\pm1/2)^2 + b^2+c^2\geq (\lambda N)^2\}$ the domain of integration. As $N\rightarrow\infty$, we can write $d_{+} = d_{-}$, $e_{+} = e_{-}$ up to terms of $O(N^{-1})$, so that
\begin{equation}
 I_2(\lambda,N) = \frac{1}{48\pi}\int\limits_{D}dadbdc\frac{c^{2s - 1}}{(r_{+}r_{-})^{2(s+1)}} + o(1),
\end{equation}
where
\begin{equation}
r_{\pm} = \sqrt{\left( a\pm\frac{1}{2} \right)^2 + b^2 + c^2}.
\end{equation}
Let us perform the inversion around $(1/2,0,0)$, so that $D$ is mapped to the region $D'$ between the sphere of radius $\lambda N + O((\lambda N)^2)$ centered around $(-1/2,0,0)$ and sphere of radius $1/(\lambda N)$ centered around $(1/2,0,0)$. The integral simplifies considerably
\begin{equation}
 I_2(\lambda,N) = \frac{1}{48\pi}\int\limits_{D'}dadbdc\frac{c^{2s - 1}}{r_{+}^{2(s+1)}} + o(1).
\end{equation}
Working up to terms vanishing as $\lambda N\rightarrow 0$, we can modify $D'$ by making the inner sphere have radius exactly $\lambda N$, and shifting the outer sphere so that it is also centered around $(-1/2,0,0)$. After these modifications, the integral becomes almost trivial, the result being
\begin{equation}
 I_2(\lambda,N) = -\frac{1}{12s}\log(\lambda N) + o(1).
\end{equation}
Combining $I_1$ and $I_2$, the dependence on $N$ drops out as expected and we find
\begin{equation}
 I(\lambda) = -\frac{1}{12 s}\log\lambda + o(1),
\end{equation}
so that $\alpha = -1/(12s)$ and $\beta = 0$.
 
\subsection{Energy operator}\label{app:eintegral}
In order to find the asymptotic behaviour of the integral \eqref{eq:energy} as $\lambda\rightarrow\infty$, which gives the one-loop properties of the energy operator, let us start by making the substitution $y_0 = r a$, $z_0 = r b$, $r_0 = r c$, after which we obtain
\begin{equation}
    G_1(x_1,x_2) = \frac{\epsilon}{6\pi}\frac{1}{r^2}\int\limits_{\mathbb{R}^3}dadbdc\frac{1}{d_+ d_- e_+ e_-}\frac{(d_++d_-)(e_++e_-)}{(d_++d_-)^{2}(e_++e_-)^{2}-(4c)^{2}},
\end{equation}
where
\begin{align*}
 d_\pm &= \sqrt{\left( a - \frac{\mu}{2} \right)^2 + b^{2} + \left( c\pm1 \right)^2}\\
 e_\pm &= \sqrt{\left( a + \frac{\mu}{2} \right)^2 + b^{2} + \left( c\pm1 \right)^2},
\end{align*}
and where we extended the domain of integration to the whole $\mathbb{R}^3$ and write $\mu = 1/\lambda$. Let us denote $J(\mu) = r^2  G_1(x_1,x_2)/\epsilon $. Analogously to the previous computation, $\mu$ acts as a point-splitting regulator for the logarithmic singularities at $(0,0,\pm1)$. We proceed along the same lines, splitting the domain into the union of the spheres of radii $N\mu$ centered at $(0,0,\pm 1)$, and the rest of $\mathbb{R}^3$, and considering the limit $N\rightarrow\infty$, $N\mu\rightarrow 0$. We start analyzing the integral $J_1(\mu,N)$ over the spheres. Concentrating on the sphere centered at $(0,0,1)$, and making the substitution $a = \mu x$, $b = \mu y$, $c = 1 +\mu z$, we find
\begin{align*}
 d_{+} = e_{+} &= 2 + \mu z + O(\mu^2)\\
 d_{-} &= \mu\sqrt{\left( x-\frac{1}{2} \right)^{2} + y^{2}+ z^{2}}+ O(\mu^2)\\
 e_{-} &= \mu\sqrt{\left( x+\frac{1}{2} \right)^{2} + y^{2}+ z^{2}}+ O(\mu^2),
\end{align*}
so that the integral becomes
\begin{equation}
 J_1(\mu,N) = \frac{1}{48\pi}\int\limits_{x^2+y^2+z^2\leq N^2}dxdydz\frac{1}{f_{+}f_{-}(f_{+} + f_{-})} + o(1),
\end{equation}
where
\begin{equation}
 f_{\pm} = \sqrt{\left( x\pm\frac{1}{2} \right)^2 + y^2 + z^2}.
\end{equation}
Notice that after scaling the variables by $1/2$, the integral is equivalent to $I_1(\lambda,2N)/4$ from \eqref{eq:I1} with $s=1/2$, so that we immediate obtain
\begin{equation}
 J_1(\mu,N) = \frac{1}{4}I_1(\mu,2N,s=1/2) + o(1) = \frac{1}{24}(\log N + \log 2 ) + o(1).
\end{equation}
Shifting to $J_2(\mu,N)$, we can use $d_{\pm} = e_{\pm}$, so that
\begin{equation}
 J_2(\mu,N) = \frac{1}{6\pi}\int\limits_{D}dadbdc\frac{1}{(d_{+}d_{-})^2}\frac{(d_+ + d_-)^2}{(d_+ + d_-)^4 - (4c)^2} + o(1),
\end{equation}
where the domain is $D=\{(a,b,c)\in\mathbb{R}^3|a^2 + b^2+(c\pm1)^2\geq (\mu N)^2\}$. Scaling the variables by $2$ and applying inversion centered at $(0,0,1)$, the integral simplifies greatly
\begin{equation}
 J_2(\mu,N) = \frac{1}{192\pi}\int\limits_{D'}dadbdc\frac{1}{\left[a^2 + b^2 + \left( c + \frac{1}{2} \right)^2\right]^{\frac{3}{2}}} + o(1),
\end{equation}
where $D'=\{(a,b,c)\in\mathbb{R}^3|a^2 + b^2 + (c\pm1/2)^2\gtrless (\mu N/2)^{\pm1}\}$. Modifying the integration domain as in the previous section, to make it into the region between two concentric spheres of mutually inverse radii, we easily find the result
\begin{equation}
 J_2(\mu,N) = -\frac{1}{24}\log\left( \frac{\mu N}{2} \right) + o(1),
\end{equation} 
so that indeed the $N$ dependence cancels in the final result and we obtain
\begin{equation}
 J(\mu) = -\frac{1}{24}\log\mu + \frac{1}{12}\log2 + o(1),
\end{equation}
so that
\begin{equation}
  G_1(x_1,x_2) =  \frac{\epsilon}{r^{2}}\left[\frac{1}{24}\log\lambda + \frac{\log 2}{12} + o(1)\right]
\end{equation}
as $\lambda\rightarrow\infty$.

\subsection{Dimensions of integer spin operators}\label{app:Sintegral}
We start the analysis of the integral \eqref{eq:integerspindim} by making the substitution $x_0 = a r/(\lambda\mu)$, $y_0 = b r/(\lambda\mu)$, $z_0 = c r/(\lambda\mu)$, after which the integral becomes
\begin{equation}
  G_{\mathrm{con.}}\left(\{x_j,s_j\}_{j=1}^4\right) = \epsilon\frac{e^{i\sum\limits_js_j\theta_j}}{r^4}(\lambda\mu)^{2(s+2)}K(\mu,\lambda),
\end{equation}
where
\begin{equation}
 K(\mu,\lambda) = -\frac{2}{3\pi}\int\limits_{\mathbb{R}^2}dadb\int\limits_0^{\infty}dc c\prod_{j=1}^{4}\frac{(4c)^{|s_j|}}{d_{j}e_{j}\left(d_{j} + e_{j}\right)^{2|s_j|}},
 \label{eq:g4integral2} 
\end{equation}
where
\begin{align*}
 d_{j} &= \sqrt{\left( a-a_j \right)^{2} + b^{2}+ \left(c - \lambda\mu \right)^{2}}\\
 e_{j} &= \sqrt{\left( a-a_j \right)^{2} + b^{2}+ \left(c + \lambda\mu \right)^{2}},
\end{align*}
where $a_1 = -1/2 - \mu/2$, $a_2 = -1/2 + \mu/2$, $a_3 = 1/2 - \mu/2$, $a_4 = 1/2 + \mu/2$. We want to study the asymptotic behaviour as $\lambda,\mu\rightarrow\infty$. Note that for fixed $\mu>0$, the integral is non-singular in the limit $\lambda\rightarrow\infty$, so we may set $\lambda = 0$, and use $d_j=e_j$. Writing $K(\mu) = K(\mu,0)$, we thus have
\begin{equation}
 K(\mu) = -\frac{2}{3\pi}\int\limits_{\mathbb{R}^2}dadb\int\limits_0^{\infty}dc \frac{c^{2s + 1}}{\prod_{j=1}^{4}d_{j}^{2|s_j|+2}},
 \label{eq:g4integral3}
\end{equation}
where
\begin{equation}
 d_{j} = \sqrt{\left( a-a_j \right)^{2} + b^{2}+ c^{2}}.
\end{equation}
As $\mu\rightarrow\infty$, the singularities collide pairwise around $(\pm1/2,0,0)$, inducing logarithmic singularities, so again, we expect $K(\mu) = \gamma\log\mu + O(1)$. The constant $\gamma$, which is related to the anomalous dimension, can be found by changing the role of $\mu$ from a point-splitting regulator, to a hard regulator, i.e. setting $\mu = 0$ in the integrand, but omitting the half-spheres of radii $\mu$ centered around $(\pm1/2,0,0)$ from the integration domain. The arguments from the previous subsections make this statement rigorous. Near $(1/2,0,0)$, we can replace $d_1 = d_2 = 1$, $d_3 = d_4$, so that the near-singularity behavior of the integral is
\begin{equation}
 K(\mu) = -\frac{8}{3}\int\limits_{\mu}\frac{dr}{r}\int\limits_{0}^{\frac{\pi}{2}}d\theta\sin\theta(\cos\theta)^{2s+1} + O(1) = \frac{4}{3(s+1)} + O(1).
\end{equation}
Hence equation \eqref{eq:g4con} follows.

\subsection{$c_{\psi\psi\bar{D}}$ OPE coefficient}\label{app:Dintegral}
In order to find the one-loop-corrected OPE coefficients $c_{\psi_{s_1}\psi_{s_2}\mathcal{O}_{s}}$, we would need to work much harder, repeating the analysis involving the auxiliary parameter $N$ on the complicated integral \eqref{eq:g4integral2}, and finding the eigenvectors of the first-order dilatation operator. Here we content ourselves with the analysis in the one-dimensional case $s_1 = s_2 = 1/2$, which gives the correction to the OPE coefficient $c_{\psi\psi\bar{D}}$. Our goal is thus to find the constant piece of \eqref{eq:g4integral3} when $|s_j|=1/2$, $s=1$
\begin{equation}
 K(\mu) = -\frac{2}{3\pi}\int\limits_{\mathbb{R}^2}dadb\int\limits_0^{\infty}dc \frac{c^{3}}{\prod_{j=1}^{4}d_{j}^{3}}.
\end{equation}
As in the previous sections, we introduce parameter $N$, and split the integration domain into the union of the two half-spheres of radii $\mu N$ centered at $(\pm1/2,0,0)$ and the rest of the 3D half-space, denoting the two resulting integrals $K_1(\mu,N)$, $K_2(\mu,N)$ respectively. $K_2(\mu,N)$ can be dealt with easily by applying the previously used methods. As $\mu N\rightarrow 0$, we can write $d_1 = d_2 = \sqrt{\left( a -1/2 \right)^2 + b^2 + c^2}$, $d_3 = d_4 = \sqrt{\left( a +1/2 \right)^2 + b^2 + c^2}$. Then, applying inversion centered at $(1/2,0,0)$ and shifting the outer half-sphere to become concentric with the inner one, we arrive at the simple integral
\begin{equation}
 K_2(\mu,N) = -\frac{2}{3\pi}\int\limits_{\mu N}^{(\mu N)^{-1}}\frac{dr}{r}\int d\Omega_{S^2_+}(\cos\theta)^3 + o(1),
\end{equation}
where $S^2_{+}$ denotes the upper half of $S^2$, i.e. $\theta\in[0,\pi/2]$. It follows that $K_2$ does not contribute to the constant term
\begin{equation}
 K_2(\mu,N) = \frac{2}{3}\log(\mu N) + o(1).
\end{equation}
Moving on to $K_1(\mu,N)$, let us focus on the half-sphere centered at $(1/2,0,0)$ and write $a = 1/2 + \mu x$, $b =\mu y$, $c =\mu z$, so that
\begin{equation}
 K_1(\mu,N) = -\frac{4}{3\pi}\int\limits_{D} dxdydz\frac{z^3}{\left[(x-\frac{1}{2})^2 + y^2 + z^2\right]^{\frac{3}{2}}\left[(x+\frac{1}{2})^2 + y^2 + z^2\right]^{\frac{3}{2}}} +o(1),
\end{equation}
where $D=\{(x,y,z)\in\mathbb{R}^3|z\geq 0, x^2 + y^2 +z^2\leq N^2\}$. The integral over $z$ can be done with the result (dropping sub-leading terms)
\begin{equation}
 K_1(\mu,N) = \frac{4}{3\pi}\int\limits_{x^2 + y^2<N^2}dxdy\left[\frac{1}{N^2} - \left( \frac{1 + \frac{x^2 + y^2}{N^2}}{r_+ + r_-} \right)^2\right] + o(1),
 \label{eq:Dope1} 
\end{equation}
where
\begin{equation}
r_{\pm} = \sqrt{\left( x\pm\frac{1}{2} \right)^2 + y^2}.
\end{equation}
The first term of \eqref{eq:Dope1} trivially integrates to $4/3$, and we will denote the remaining integral $L(N)$. Scaling the variables by $N$, we find
\begin{equation}
 L(N) = -\frac{4}{3\pi}\int\limits_{x^2 + y^2<1}dxdy\left( \frac{1 + x^2 + y^2}{\sqrt{\left( x+\frac{1}{2N} \right)^2 + y^2} + \sqrt{\left( x-\frac{1}{2N} \right)^2 + y^2}} \right)^2.
\end{equation}
We can simplify the integral by repeating our trick of splitting the integration domain into the disc of radius $M/N$ and the remaining annulus and consider the limit $M\rightarrow\infty$, $M/N\rightarrow 0$. Denote the disc integral by $L_1(M,N)$ and the annulus integral by $L_2(M,N)$. When evaluating $L_2$, we can set $1/N = 0$ in the integrand and arrive at
\begin{equation}
 L_2(M,N) = -\frac{2}{3}\int\limits_{\frac{M}{N}}^{1}drr\frac{(1+r^2)^2}{r^2} + o(1) = \frac{2}{3}\log\left( \frac{M}{N} \right) - \frac{5}{6} + o(1),
\end{equation} 
so that $L_2$ contributes $-5/6$ to the constant term. It remains to find the constant term in $L_1(M,N)$. Scaling the variables by $N$, the integral becomes
\begin{equation}
 L_1(M,N) =  -\frac{4}{3\pi}\int\limits_{x^2 + y^2<M}dxdy\frac{1}{\left(\sqrt{\left( x+\frac{1}{2} \right)^2 + y^2} + \sqrt{\left( x-\frac{1}{2} \right)^2 + y^2}\right)^2} + o(1).
\end{equation}
The angular integration can be done in terms of elliptic integrals or hypergeometric functions, and the radial integral can then be expanded as $M\rightarrow\infty$
\begin{equation}
 L_1(M,N) = -\frac{2}{3}\log M + \frac{1 - 4\log 2}{3} + o(1).
\end{equation}
Hence, putting all the constants together
\begin{equation}
 K_1(\mu,N) = -\frac{2}{3}\log N - \frac{8\log2 - 5}{6} + o(1),
\end{equation}
and so
\begin{equation}
 K(\mu) = \frac{2}{3}\log\mu - \frac{8\log2 - 5}{6} + o(1)
\end{equation}
as $\mu\rightarrow 0$.

\section{The four-point function without the defect}\label{app:nodefect}
In this appendix, we will compute the one-loop correction to the four-point function of $\phi$ in the $\phi^4$ theory without the defect in order to find properties of $\epsilon$ at one loop. Placing the four insertions on a line, with distances $|x_{12}| = |x_{34}| = r$, $|x_{13}| = |x_{24}| = r/\mu$, $\mu\ll 1$, the OPE predicts
\begin{equation}
  G(x_1,x_2,x_3,x_4) = \frac{1}{r^{4\Delta_{\sigma}}}\left[1 + c_{\sigma\sigma\epsilon}^2\mu^{2\Delta_\epsilon}(1+o(1))\right].
 \label{eq:openodefect} 
\end{equation}
The free-theory values are $\Delta_{\sigma} = 1-\epsilon/2$, $\Delta_\epsilon = 2 - \epsilon$, $c_{\sigma\sigma\epsilon} = \sqrt{2}$, where the first result holds also at the Wilson-Fisher fixed point up to corrections of $O(\epsilon^2)$. The one-loop self-energy vanishes in the massless $\phi^4$ theory, so the only contribution comes from the contact four point interaction
\begin{equation}
  G_1 = -(2\pi)^4 g\int d^4x_0G(x_1,x_0)G(x_2,x_0)G(x_3,x_0)G(x_4,x_0) = 
 -\frac{\epsilon}{3\pi^2}\int d^4x_0\frac{1}{x_{01}^2x_{02}^2x_{03}^2x_{04}^2}.
\end{equation}
After scaling the integration variables as $x_0 = r y/\mu$, the integral becomes
\begin{equation}
  G_1 = \frac{\epsilon}{r^4}\mu^4 H(\mu),
\end{equation}
where
\begin{equation}
 H(\mu) = -\frac{1}{3\pi^2}\int d^4y\frac{1}{\left(y - \frac{1+\mu}{2}\hat{n}\right)^2\left(y - \frac{1-\mu}{2}\hat{n}\right)^2\left(y + \frac{1-\mu}{2}\hat{n}\right)^2\left(y + \frac{1+\mu}{2}\hat{n}\right)^2},
\end{equation}
where $\hat{n}$ is a unit vector in a fixed direction. As $\mu\rightarrow0$, the integral develops logarithmic singularities at $y=\pm\hat{n}/2$. We can compute the logarithmic and constant piece exactly as before, splitting $H(\mu) = H_1(\mu,N) + H_2(\mu,N)$, where $H_1(\mu,N)$ is the integral over the union of the two spheres of radii $\mu N$, centered at $\pm\hat{n}/2$, and $H_2(\mu,N)$ over their complement in $\mathbb{R}^4$. Again, we work in the limit $N\rightarrow\infty$, $\mu N\rightarrow 0$. Working with $H_1$ and focusing on the sphere centered at $\hat{n}/2$, we can set the distances to the far singularities equal to one, and upon rescaling the variables by $\mu$, we obtain
\begin{equation}
 H_1(\mu,N) = -\frac{2}{3\pi^2}\int\limits_{R^2<N^2}dy^4\frac{1}{\left(y - \frac{1}{2}\hat{n}\right)^2\left(y + \frac{1}{2}\hat{n}\right)^2} + o(1).
\end{equation}
The integral can be evaluated exactly
\begin{equation}
 \frac{1}{\pi^2}\int\limits_{R^2<N^2}dy^4\frac{1}{\left(y - \frac{1}{2}\hat{n}\right)^2\left(y + \frac{1}{2}\hat{n}\right)^2} \stackrel{N>\frac{1}{2}}{=} \log\left( 4N^2 + 1 \right) + 1,
\end{equation}
giving the following asymptotics
\begin{equation}
 H_1(\mu,N) = -\frac{4}{3}\log N - \frac{2}{3} + o(1).
\end{equation}
Moving on to $H_2(\mu,N)$, we can set $\mu = 0$ in the integrand, and use our usual trick of doing the inversion centered at $\hat{n}/2$, after which the integration domain becomes the region between the sphere of radius $\mu N + O((\mu N)^2)$ centered at $-\hat{n}/2$ and the sphere of radius $1/(\mu N)$ centered at $\hat{n}/2$. Ignoring terms vanishing as $\mu N\rightarrow 0$, we can make the former radius exactly $\mu N$ and make the spheres concentric, so that
\begin{equation}
 H_2(\mu,N) = -\frac{1}{3\pi^2}\int\limits_{\mu N<R<(\mu N)^{-1}}\frac{d^4y}{R^4} + o(1) = \frac{4}{3}\log(\mu N) + o(1).
\end{equation}
Hence
\begin{equation}
 H(\mu) = \frac{4}{3}\log\mu - \frac{2}{3} + o(1).
\end{equation} 
Comparing with the expansion \eqref{eq:openodefect}, this gives the following properties of the energy operator
\begin{align}
 \Delta_\epsilon &= 2 - \frac{2}{3}\epsilon + O(\epsilon^2)\\
 c_{\sigma\sigma\epsilon} &= \sqrt{2}\left( 1 - \frac{\epsilon}{6} \right) + O(\epsilon^2).
\end{align}

\bibliographystyle{JHEP}
\bibliography{CFT}

\end{document}